\documentclass[aps,a4,floatfix,nofootinbib]{revtex4}  

\usepackage{fmtcount} 
\usepackage{graphicx}
\usepackage{float}
\usepackage{psfrag}
\usepackage{fixmath}
\usepackage{amsmath, amsfonts, amssymb, amsthm, amscd} 
\usepackage{marvosym} 
\usepackage{yfonts}  
\DeclareGraphicsExtensions{ps,eps,ps.gz,frh.eps,ml.eps}
\usepackage{amstext,amssymb,amsbsy}
\usepackage{braket}
\usepackage{color}
\usepackage{relsize}
\usepackage{mathrsfs} 
\usepackage{mathcomp} 
\usepackage{wasysym} 
\usepackage{upgreek}
\usepackage[all]{xy}
\usepackage{ dsfont }
\usepackage{ stmaryrd }
\usepackage[nottoc,numbib]{tocbibind}
\usepackage{array}

\usepackage{url}
\usepackage{hyperref}

\def\XXint#1#2#3{{\setbox0=\hbox{$#1{#2#3}{\int}$ }
\vcenter{\hbox{$#2#3$ }}\kern-.6\wd0}}

\makeatletter

\newcommand{\Rmnum}[1]{\expandafter\@slowromancap\romannumeral #1@}
\makeatother

\makeatletter

\def\env@blscases{
  \let\@ifnextchar\new@ifnextchar
  \left.
  \def\arraystretch{1.2}
  \array{@{}l@{\quad}l@{}}
}
\makeatother

\makeatletter

\def\env@rcases{
  \let\@ifnextchar\new@ifnextchar
  \left.
  \def\arraystretch{1.2}
  \array{@{}l@{\quad}l@{}}
}
\makeatother

\DeclareMathOperator{\e}{\operatorname{e}}

\DeclareMathOperator{\tr}{\operatorname{tr}}

\newcommand{\abs}[1]{\lvert#1\rvert}

\begin {document}
\title{Nuclear thermodynamics from chiral low-momentum interactions}
\author{Corbinian\ Wellenhofer$^{1}$}
\author{Jeremy\ W.\ Holt$^{2}$}
\author{Norbert\ Kaiser$^{1}$}
\author{Wolfram\ Weise$^{1,3}$}
\affiliation{\vspace*{0.5mm} $^1$Physik Department, Technische Universit\"{a}t M\"{u}nchen, D-85747 Garching, Germany \\ 
$^2$Department of Physics, University of Washington, Seattle, Washington 98195, USA \\  
$^3$ECT$^{\, *}$, Villa Tambosi, I-38123 Villazzano (TN), Italy}
\date{27 June 2014}

\begin{abstract}
\noindent
We investigate the thermodynamic equation of state of isospin-symmetric nuclear matter with microscopic 
nuclear forces derived within the framework of chiral effective field theory. Two- and 
three-body nuclear interactions constructed at low resolution scales form the basis for a perturbative
calculation of the finite-temperature equation of state. 
The nuclear force models and many-body 
methods are benchmarked against bulk properties of isospin-symmetric nuclear matter at zero temperature, 
which are found to be well reproduced when chiral nuclear interactions constructed at the lowest resolution 
scales are employed.
The calculations are then extended to finite temperatures, where we focus on 
the liquid-gas phase transition and the associated critical point. 
The Maxwell construction is 
applied to construct the physical equation of state, and the value of the critical temperature is 
determined to be $T_c =17.2-19.1$\,MeV, in good agreement with the value extracted from 
multifragmentation reactions of heavy ions.
\end{abstract}
\maketitle

\section{Introduction}
The equation of state (EoS) of nuclear matter is of fundamental importance for heavy-ion collisions and for a range of 
astrophysical phenomena, including neutron star structure and evolution, nucleosynthesis, 
as well as the dynamics of core-collapse supernovae and binary neutron star mergers. The recent
observation of two-solar-mass neutron stars \cite{Demorest:2010bx, Antoniadis:2013pzd} 
places strong constraints on the neutron matter EoS. To support neutron stars of 
such mass, the EoS has to be comparatively stiff, which at first glance appears to favor
neutron star models with primarily nucleonic degrees of freedom and challenges models
that include exotic condensates or deconfined quark matter \cite{Demorest:2010bx, Lattimer:2012nd, Hell:2014xva}. 
The interpretation of expected observations of gravitational waveforms linked to binary neutron
star (or neutron star---black hole) mergers provides further motivation for calculations of the dense
nucleonic matter equation of state with reliable uncertainty estimates.
Such astrophysical applications require a realistic
EoS for neutron matter with a small admixture of protons. As a prerequisite for any such 
discussion, an essential condition is to have an EoS for isospin-symmetric nuclear 
matter that is consistent with empirical constraints provided by nuclear thermodynamics.
The present work focuses on this issue.

With the development of chiral effective field theory ($\chi$EFT), high-precision two- and many-nucleon
forces constrained by the symmetry breaking pattern of QCD provide the foundation for systematic studies of low-energy
nuclear structure and reactions. Once the low-energy constants that parametrize unresolved short-distance 
nuclear dynamics are fixed by fits to 
few-nucleon observables (as a function of the chosen resolution scale), nuclear many-body properties result as 
pure predictions. Empirical properties of infinite homogeneous nuclear matter, such as the 
saturation point and compressibility of isospin-symmetric nuclear matter at zero temperature as well as the 
first-order transition from a liquid to a vapor phase at finite temperature, are then nontrivial tests of the 
many-body methods and nuclear force models. Particularly the critical point of the liquid-gas phase transition is essential in constraining the finite-temperature domain of the nuclear EoS. 
Estimates for this point have been obtained through the analysis of data from multifragmentation, fission 
and compound nuclear decay experiments \cite{Elliott:2013pna, Karnaukhov:2008be}. The critical 
temperature $T_c$ in particular was located at approximately $18 \, \text{MeV}$.  

In recent years, various aspects of the zero-temperature nuclear EoS from $\chi$EFT have been 
studied in detail by numerous authors within widely different many-body frameworks 
\cite{Kaiser:2001jx,Hebeler:2009iv,Hebeler:2010xb,Tews:2012fj,HoltJW13a,Coraggio13,Kruger:2013kua,
Carbone2013,Gezerlis13,Holt:2013fwa,Hagen2014,Drischler14,Coraggio:2014nvaa,Roggero14,Wlazlowski2014}.
The picture that arises is that low-momentum microscopic nuclear interactions associated with a 
resolution scale around $400-450$ MeV facilitate the convergence of the EoS in many-body 
perturbation theory at and near nuclear matter saturation density ($\rho_0 \simeq 0.17$\,fm$^{-3}$), 
while an accurate treatment of very low-density matter must account for nuclear clustering and 
nonperturbative features of the nucleon-nucleon interactions associated with the physics of large 
scattering lengths. The thermodynamic properties of neutron matter have been studied in Ref.\
\cite{Tolos:2007bh}, while in-medium chiral perturbation theory at finite temperature was
used to explore isospin-asymmetric nuclear matter over a range of proton fractions and densities 
\cite{Fiorilla:2011sr}. 
For an earlier study of nuclear matter at finite temperatures using phenomenological Skyrme
forces, see
Ref. \cite{mosel}.
In the present work we take the initial steps toward a complementary microscopic
study of nuclear thermodynamics across the densities and isospin asymmetries relevant for 
astrophysical simulations of supernovae and neutron stars. We compute the thermodynamic equation 
of state of isospin-symmetric nuclear matter from several sets of chiral low-momentum interactions
and investigate their thermodynamic consistency. 

The paper is organized as follows. 
In Sec. \ref{sec1} we provide details of the different two- and three-body potentials that specify 
the microscopic input for the subsequent calculations. In Sec. \ref{sec2} we proceed with a 
discussion of many-body perturbation theory generalized to finite temperatures. The use of a 
temperature- and density-dependent effective two-nucleon potential  \cite{Holt:2008,Holt:2009ty,Hebeler:2009iv,
Hebeler:2010xb} to approximate the second-order 
three-body contributions is discussed in detail, as well as the temperature-dependent self-energy
corrections to the single-nucleon energies. In Sec. \ref{sec3} we examine our results 
for the thermodynamic EoS of isospin-symmetric nuclear matter. Finally, in Sec. \ref{sec5} 
we give a summary of the main results and an outlook for future investigations of the nuclear 
many-body problem with the use of chiral low-momentum interactions.
%
%
%
\section{Chiral low-momentum two- and three-body interactions} \label{sec1}
The modern theory of nuclear forces is based on chiral effective field theory \cite{Epelbaum:2008ga,
Machleidt:2011zz}, the low-energy realization of quantum chromodynamics. Its regime 
of applicability is governed by a separation of scales, where the hard scale is given by the chiral
symmetry breaking scale $\Lambda_\chi \sim 1$\,GeV, and the soft scale is associated with small nucleon
momenta $Q$ that for many phenomena of interest are of the same order of magnitude as the pion 
mass $m_\uppi$. The Feynman diagrams contributing to interactions between nucleons are then 
organized in an expansion in powers of the parameter $Q/\Lambda_\chi$. Short-distance dynamics 
associated with the length scale $1/\Lambda_{\chi}$ is parametrized by low-energy constants 
(LECs) that are generally fixed by fitting to two-nucleon scattering phase shifts and in the case of 
nuclear three-body forces to properties of $^3$H and $^3$He. The results of these fitting procedures are 
not unique, and there exist various sets of LECs in the literature (e.g., \cite{Buettiker:1999ap,Epelbaum:2002ji,Machleidt:2011zz,Rentmeester:2003mf}), 
all of which lead (by construction) to consistent results in the few-body sector.

Because of the limited energy regime accessible to $\chi$EFT, chiral nuclear interactions are typically
regulated at a scale $\Lambda$ lying between the low- and high-energy regimes: $Q < \Lambda <
\Lambda_\chi$. Regarding two-body forces, a common way to enforce such a restriction is to multiply the nucleon-nucleon potential $V_{\text{NN}}$ with a 
smooth regulator function of the form 
\begin{align} \label{EMregulator}
f(p,p')=\exp \left[-(p/\Lambda)^{2n}-(p'/\Lambda)^{2n} \right],
\end{align} 
where $p$ and $p'$ are the absolute values of the relative momenta of the two nucleons 
before and after the collision\footnote[1]{To be precise, $p$ (and similarly $p'$) is defined as half of the 
relative momentum of the two nucleons, i.e., $p=|\vec k_1-\vec k_2|/2$.}. 
In the following, we employ two-nucleon potentials constructed at the resolution scales 
$\Lambda =414, 450, 500\,\text{MeV}$ (see Refs.\ \cite{Entem:2003ft,Coraggio07b,Coraggio13,
Coraggio:2014nvaa} for additional details). In each case, we employ as well the respective 
next-to-next-to-leading order (N2LO) chiral three-body interaction (depending on the parameters 
$c_E$, $c_D$ and $c_{1,3,4}$). The implementation of 
consistent
N3LO chiral many-nucleon forces remains a challenge in contemporary nuclear structure theory, but 
progress toward this end is being achieved \cite{Tews:2012fj}. We hereafter denote these three sets 
of chiral two- and three-body potentials by n3lo414, n3lo450 and n3lo500. Because of the different regulating 
functions used in the respective potentials, different values of LECs emerge from fits to few-body 
observables. The resulting values for the five LECs that appear in the leading-order three-body diagrams 
are given in the first three rows of \mbox{Table \ref{table:LECs}}.

As the cutoff scale is reduced below 500\,MeV in the construction of chiral nuclear interactions, precision 
fits to nucleon-nucleon scattering phase shifts deteriorate \cite{Bogner:2009bt,Marji:2013uia}. An alternative scheme for obtaining 
low-momentum nuclear interactions is to employ renormalization group (RG) techniques \cite{Bogner:2009bt,Furnstahl:2012fn} that by construction leave low-energy observables invariant. In the case of an evolution of the
NN potential based on half-on-shell $K$-matrix equivalence the resulting potential is usually denoted 
by $V_{\text{low-}k}(\Lambda)$, with $\Lambda$ being a sharp cutoff in momentum space 
\cite{Bogner:2003wn}. For cutoffs in the range $\Lambda\simeq 2.1\,\text{fm}^{-1}$, the RG evolved potential 
is universal, i.e., independent of the input potential. This method has the advantage of 
producing low-momentum NN potentials directly through the evolution of partial-wave matrix elements, 
however, the inclusion of induced many-nucleon forces is crucial. In view of this, Nogga \emph{et al.}\ 
\cite{Nogga:2004ab} have used the leading-order 3N forces with the values of the $c_{1,3,4}$ constants 
equal to the ones extracted by the Nijmegen group in an analysis of NN scattering data 
\cite{Rentmeester:2003mf} and determined $c_E$ and $c_D$ by fitting to the binding energies of 
$^3$H, $^3$He and $^4$He. The resulting LECs for two different $V_{\text{low-}k}$ potentials 
(both constructed by evolving the n3lo500 NN potential) can be found in the last two rows of 
Table \ref{table:LECs}.  
\begin{table}[H]
\begin{center}
\begin{minipage}{14cm}
\setlength{\extrarowheight}{1.5pt}
\hspace*{18mm}
\begin{tabular}{|l|c|c|| c |c |c |c |c |}
\hline 
 & $\Lambda\,[\text{fm}^{-1}]$ &$n$ &$c_E$ & $c_D$ & $c_1 \,[\text{GeV}^{-1}]$ &$c_3 \,[\text{GeV}^{-1}]$ 
&$c_4 \,[\text{GeV}^{-1}]$  \\ \hline   \hline 
 n3lo500   &$2.5$& $2$&-0.205 &-0.20   &-0.81  &-3.2      &5.4   \\
 \hline 
n3lo450   &$2.3$& $3$&-0.106 &-0.24   &-0.81  &-3.4      &3.4   \\
n3lo414   &$2.1$&$10$&-0.072 &-0.4   &-0.81  &-3.0      &3.4   \\
 \hline 
VLK23  &$2.3$& $\infty$ &-0.822 &-2.785  &-0.76  &-4.78     &3.96  \\
VLK21  &$2.1$& $\infty$ &-0.625 &-2.062  &-0.76  &-4.78     &3.96  \\
 \hline 
\end{tabular}
\end{minipage}
{\vspace{0mm}}
\begin{minipage}{14cm}
\caption
{The different sets of chiral low-momentum two- and three-body interactions used in this work. For the n3lo NN potentials relative momenta are restricted by a smooth regulator with cutoff scale 
$\Lambda$ and steepness parameter $n$, whereas in the case of the VLK two-body potentials there is a sharp cutoff. With 
the cutoff scale and the regulator width taken from the respective two-body regulator the different three-body potentials are 
completely determined by the values of $c_E$, $c_D$ and $c_{1,3,4}$.}
\label{table:LECs}
\end{minipage}
\end{center}
\end{table}

To summarize, we will analyze nuclear thermodynamics through five different sets of two- and 
three-body potentials. These can be used to probe a variety of aspects associated with the 
choice of resolution scale and low-energy constants. Of particular interest will be the comparison of nuclear potentials defined
at the same resolution scale but constructed via RG methods or by refitting LECs. 
\vspace*{25mm}
\section{Many-body perturbation theory for nuclear matter} \label{sec2}

With the use of low-momentum interactions, many-body perturbation theory (MBPT) becomes applicable 
for the investigation of the nuclear many-body system \cite{Bogner:2005sn}. In the present section we 
recall the main aspects of this framework, and give analytical expressions for the different terms 
contributing to the free energy density, both in the case of zero as well as finite temperatures. We begin by
summarizing the main results for a free Fermi gas. The interacting many-nucleon system is then introduced 
and we present the general perturbation series for the energy density (zero temperature) and the 
grand canonical potential density (finite temperature), and discuss the relationship between both series, which
motivates the Kohn-Luttinger-Ward formalism. Explicit expressions for the different contributions at first and 
second order in MBPT are then given in terms of partial-wave amplitudes. Following this, we compute the 
temperature- and density-dependent effective NN potential from the leading-order chiral three-nucleon
force. We then examine the anomalous contributions which arise in the case of finite temperatures, and finally 
we calculate the temperature-dependent self-energy corrections to the single-nucleon energies.

\subsection{Free Fermi gas} \label{sec21}
Although the free Fermi gas can easily be treated fully relativistically, we give here the expressions for the
energy density and grand canonical potential density for a nonrelativistic Fermi gas and include relativistic 
effects by a correction term. The single-particle energies are given by the formula $\varepsilon_p=p^2/2M$. 
\subparagraph{Zero temperature.} The particle density of the system in the ground state depends only on the Fermi momentum $k_F$, which signifies the highest occupied energy level. The density is given by
\begin{align} \label{rhoT0}
\rho(k_F)=\frac{2 k_F^3}{3 \pi^2}.
\end{align}
This expression is exact both for a relativistic and a nonrelativistic Fermi gas. The energy per particle, including the 
first relativistic correction, is given by
\begin{align} \label{energytzero}
\bar{E}_0(k_F)=\frac{3 k_F^2}{10 M}-\frac{3 k_F^4}{56 M^3}.
\end{align}

\subparagraph{Finite temperatures.}
The grand canonical potential density (with the relativistic correction constructed in Ref.\ \cite{Fritsch:2002hp}) is given by 
\begin{align} \label{OmegaNorbert}
\Omega_0(\mu_0,T)=
-\frac{2}{3 \pi^2}\int \limits_0^{\infty} \! \mathrm{d}p \, \frac{p^4}{M} n_p\,\,
-\,\, \frac{1}{4 \pi^2}\int \limits_0^{\infty} \! \mathrm{d}p \, \frac{p^6}{M^3} n_p ,
\end{align}
where $\mu_0$ is the (nonrelativistic) chemical potential, $\beta=1/T$ is the inverse temperature and 
$n_p=1/[1+\exp(\beta(\varepsilon_p-\mu_0))]$ is the Fermi-Dirac distribution function. 
The particle density then follows from a standard thermodynamic relation, i.e.,
\begin{align} \label{rhoformulae}
\rho(\mu_0,T) =-\frac{\partial\Omega_0}{\partial \mu_0} =
\frac{2}{\pi^2} \int \limits_0^{\infty}\! \mathrm{d}p \, p^2  \, n_p =
-\sqrt{2}\left(\frac{M}{\beta \pi}\right)^{\frac{3}{2}} \text{Li}_{3/2}\big(-\exp(\beta \mu_0)\big) ,
\end{align}
where $\text{Li}_\nu(x)=\sum^{\infty}_{k=1} k^{-\nu} x^k$ is the poly-logarithmic function of index $\nu$.
Because for the free Fermi gas the pressure isotherms $P_0(\mu_0)=-\Omega_0(\mu_0)$ are strictly convex, the above relation is 
invertible with respect to $\mu_0(\rho,T)$. For densities $\rho < -\sqrt{2}\left[M/(\beta \pi)\right]^{3/2} \text{Li}_{3/2}(-1)\simeq 
0.000727\,(T/\text{MeV})^{3/2} \,\text{fm}^{-3}$ the chemical potential is negative, and its behavior in the limit of vanishing 
densities is given by
\begin{align} \label{mufreelimit}
\mu_0(\rho,T) \xrightarrow{\rho \rightarrow 0} \frac{1}{\beta} \ln \left( \frac{\rho}{\sqrt{2}} \left( \frac{\beta \pi}{M} \right)^{\frac{3}{2}}
 \right) ,
\end{align}
which, using $\text{Li}_\nu(x) \xrightarrow{x\rightarrow 0} x$, follows from inverting Eq.\ (\ref{rhoformulae}) in the limit
 $\mu_0 \rightarrow -\infty$.

From Eqs.\ (\ref{OmegaNorbert}) and (\ref{rhoformulae}) one can calculate the free energy density, $F_0(\mu_0,T)=\mu_0 
\rho(\mu_0,T)+\Omega_0(\mu_0,T)$, which reproduces the energy density $E_0(k_F)=\rho(k_F) \bar E_0(k_F)$ 
in the zero-temperature limit:
\begin{align}
F_0(\mu_0,T)\xrightarrow{T\rightarrow 0} E_0(k_F)\;\Big|_{\rho\, \text{fixed}} \;,\;\text{where}\; \mu_0 \xrightarrow{T\rightarrow 0} 
\frac{k_F^2}{2M} \;\Big|_{\rho\, \text{fixed}} .
\end{align}

\subsection{Many-body perturbation series: general discussion} \label{sec22}

\subparagraph{Zero temperature.} The energy density $E=\braket{\Psi_0|\mathcal{H}|\Psi_0}/\braket{\Psi_0|\Psi_0}$ of the 
interacting many-nucleon system with Hamiltonian $\mathcal{H}=\mathcal{H}_0+\lambda \mathcal{V}$ is given by the following 
perturbation series (known as the Brueckner-Goldstone formula \cite{Brueckner:1958zz,Goldstone:1957zz}):
\begin{align} \label{BG-series}
E(k_F)=E_0(k_F)+\lambda E_{1}(k_F)+\lambda ^2E_{2}(k_F)+\mathcal{O}(\lambda^3),
\end{align} 
where $\Ket{\Psi_0}$ is the exact ground state of the interacting system, $\mathcal{H}_0$ is the Hamiltonian of the 
non-interacting system, $\mathcal{V}=\mathcal{V}_{\text{NN}}+
\mathcal{V}_{\text{3N}}$ is the interaction part of the Hamiltonian, and $\lambda$ is a counting parameter (introduced only 
for book keeping reasons). In Eq.\ (\ref{BG-series}), $E_0(k_F)$ corresponds to the energy density of a non-interacting 
nucleon gas. The different contributions contained in $E_1$ and $E_2$ are given in terms of expectation values with respect 
to the non-interacting ground state $\Ket{\Phi_0}$, which is characterized by the occupation of all energy levels below the 
Fermi energy $\varepsilon_F=k_F^2/2 M$ (where $M\simeq 938.9\,\text{MeV}$ is the average nucleon mass). Hence, all terms in Eq.\ (\ref{BG-series}) are parametrized by $k_F$, which is related 
to the nucleon density via Eq.\ (\ref{rhoT0}). From the perspective of statistical mechanics Eq.\ (\ref{BG-series}) therefore 
amounts to a calculation in the canonical ensemble, with the free energy density given by $F(\rho,T=0)=E(k_F)$.

\subparagraph{Finite temperatures.}
The imaginary-time (Matsubara) formalism leads to the following perturbation series for the grand canonical potential density $\Omega$, 
or the negative pressure $P$:
\begin{align} \label{MBPTfinT}
\Omega(\mu,T)=-P(\mu,T)=\Omega_0(\mu,T)+\lambda \Omega_{1}(\mu,T)+\lambda^2 \Omega_{2}(\mu,T)+\mathcal{O}(\lambda^3).
\end{align}
In contrast to the zero-temperature perturbation series, all terms in the above expression are functions of the chemical potential $\mu$ 
of the interacting system (in addition to temperature $T$), corresponding to the grand canonical ensemble. Moreover, compared to 
Eq.\ (\ref{BG-series}) there are additional terms (beginning at order $\lambda^2$), the so-called \textit{anomalous contributions}. 
Apart from these differences, the explicit form of the different terms in Eq.\ (\ref{MBPTfinT}) is the same as in the zero-temperature case, 
except one has finite-temperature Fermi-Dirac distributions instead of step functions. 

\subparagraph{Kohn-Luttinger-Ward formalism.}
The zero-temperature limit of the thermodynamic equation of state calculated using Eq.\ (\ref{MBPTfinT}) does not reproduce the 
equation of state obtained from the Brueckner-Goldstone formula. In the case of realistic two- and three-body forces the 
Brueckner-Goldstone formula is known to produce the desired van der Waals type EoS of isospin-symmetric nuclear matter. This 
result cannot be obtained from Eq.\ (\ref{MBPTfinT}) because in the part of the liquid-gas coexistence region where the (analytical) free energy 
density $F(\rho,T)$ is nonconvex (with respect to $\rho$), the corresponding pressure isotherms (as functions of chemical potential) $P(\mu,T)$ are 
multivalued. It is impossible to obtain such a feature in a grand canonical calculation; or more generally, in the case of a system that 
is unstable with respect to phase mixing, the canonical and the grand canonical descriptions are not equivalent, and the Legendre 
transformation between both is not invertible \cite{legendre}.

For a consistent continuation of the Brueckner-Goldstone formula to finite temperatures, one should use a perturbation series for the 
free energy density $F(\rho,T)$, i.e., employ a calculation in the canonical ensemble. The method for constructing such a 
perturbation series for $F(\rho,T)$, based on the grand canonical one, Eq.\ (\ref{MBPTfinT}), was introduced by Kohn and 
Luttinger \cite{Kohn:1960zz} and elaborated by Luttinger and Ward \cite{Luttinger:1960ua}. This method, which was used also by other 
authors (e.g., \cite{Fritsch:2002hp,Tolos:2007bh,Fiorilla:2011sr}), works as follows.
Instead of calculating $F(\rho,T)$ directly from Eq.\ (\ref{MBPTfinT}) via $\rho=-\partial \Omega / \partial \mu$ and $F=\Omega+\mu\rho $, 
one constructs an expansion about the non-interacting system (free Fermi gas) with (formally) the same density:
\begin{align} \label{KLWeq1}
\rho(\mu_0,T)=-\frac{\partial \Omega_0(\mu_0,T)}{\partial \mu_0} \equiv
-\frac{\partial \Omega(\mu,T)}{\partial \mu}=\rho(\mu,T) .
\end{align}
The chemical potential is then formally expanded in terms of the counting parameter, $\mu=\mu_0+\lambda \mu_1+\lambda^2 \mu_2+
\mathcal{O}(\lambda^3)$. Expanding each term $\partial\Omega_i(\mu,T)/ \partial \mu$ around $\mu_0$ and solving Eq.\ (\ref{KLWeq1}) 
iteratively for increasing powers of $\lambda$ gives expressions for the $\mu_i$, $i \geq 1$, as functions of $\mu_0$, e.g.,
\begin{align}
\mu_1(\mu_0,T)=-\frac{\partial \Omega_1 / \partial \mu}{\partial^2 \Omega_0 / \partial \mu^2} \bigg|_{\mu_0} .
\end{align}
Finally, expanding each term $\Omega_i(\mu,T)$ in the defining relation for the free energy density, $F=\Omega+\mu \rho$, around 
$\mu_0$ leads to the following expression for $F(\mu_0,T)$:
\begin{align} \label{KLWseries}
F(\mu_0,T)=F_0(\mu_0,T)+\lambda \Omega_1(\mu_0,T)+\lambda^2 \left(\Omega_2 (\mu_0,T)-
\frac{1}{2} \frac{\left(\partial \Omega_1 / \partial \mu_0\right)^2}{\partial^2 \Omega_0 / \partial \mu_0^2} \right)+\mathcal{O}(\lambda^3).
\end{align}
Note that all terms in this equation are evaluated at $\mu_0$, which is in one-to-one correspondence with the nucleon density $\rho$ as 
specified by Eq.\ (\ref{rhoformulae}). Moreover, it can be verified that for spherically symmetric Fermi surfaces and rotationally invariant 
as well as isospin-symmetric interactions the additional derivative term at order $\lambda^2$, which is hereafter referred to as the 
second-order \textit{anomalous derivative term} (ADT), cancels the second-order anomalous contribution in the zero-temperature limit. 
Therefore, the above expression for the free energy density satisfies the desired consistency relation
\begin{align}
F(\mu_0,T)\xrightarrow{T\rightarrow 0} E(k_F)\;\Big|_{\rho\, \text{fixed}} \;,\;\text{where}\; \mu_0 \xrightarrow{T\rightarrow 0}  
\frac{k_F^2}{2M}\;\Big|_{\rho\, \text{fixed}}  .
\end{align}
The EoS obtained with Eq.\ (\ref{KLWseries}) can then of course not be the same as the one resulting from the grand canonical 
expression, Eq.\ (\ref{MBPTfinT}). The deviations between them are from the truncation of the Taylor expansions of 
$\Omega(\mu,T)$ and $\partial \Omega(\mu,T)/\partial \mu$ around $\mu_0$ at order $\lambda^2$.

\subsection{Many-body perturbation series: contributions}  \label{sec23}
\begin{figure}[H]
\vspace*{-1.9cm}
\hspace*{-0.1cm}
\includegraphics[width=1.05\textwidth]{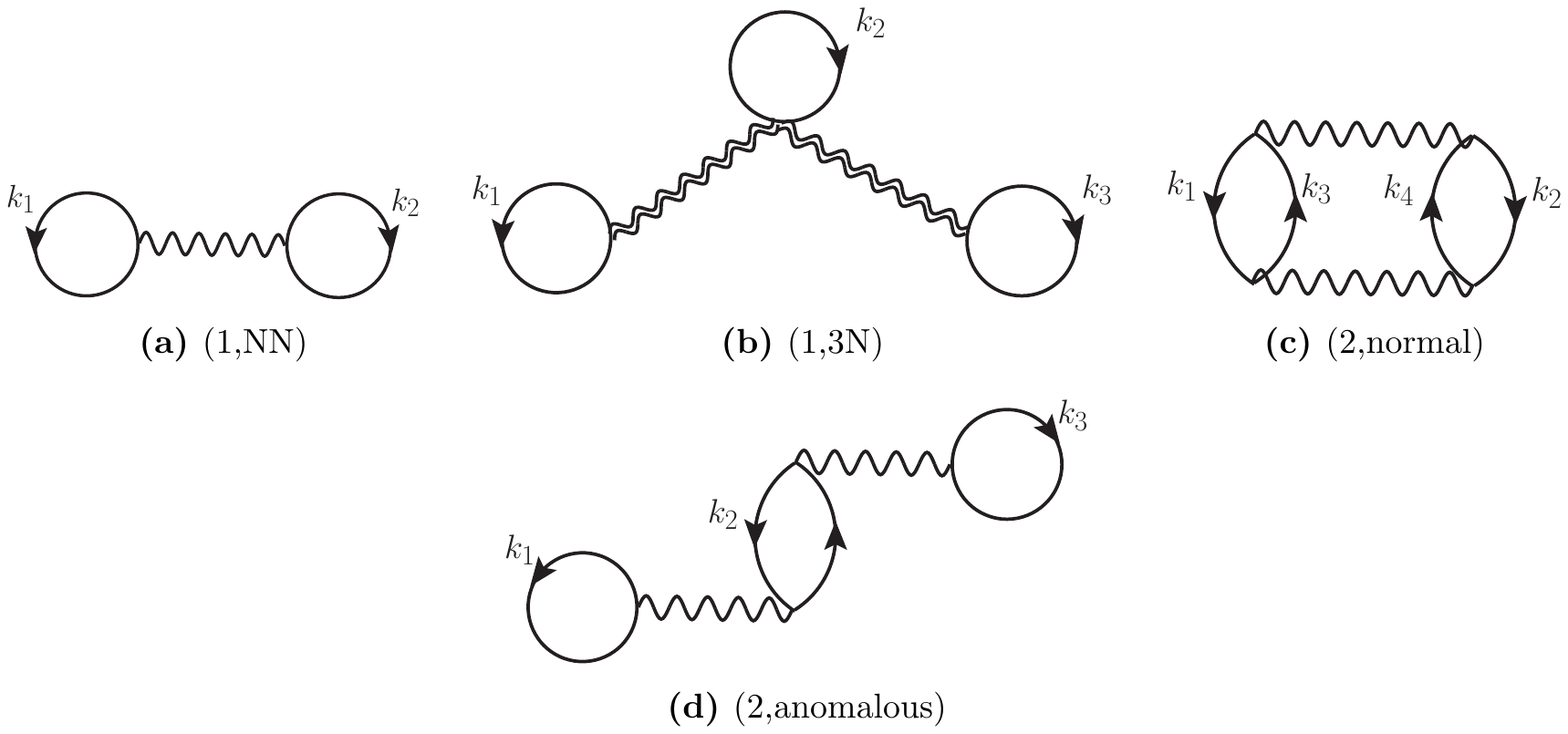} 
\vspace*{-14.9cm}
\caption{Antisymmetrized Goldstone diagrams representing the (a) first-order NN, (b) first-order 3N, (c) second-order normal NN,
and (d) second-order anomalous NN contributions. Wavy lines represent interactions 
mediated by $\tilde{V}_{\text{NN}}$; double-wavy lines symbolize $\tilde{V}_{\text{3N}}$. }
\label{fig_mbds}
\end{figure}

We now give the explicit form of the first- and second-order contributions in the Kohn-Luttinger-Ward formula, Eq.\ (\ref{KLWseries}). 
For reasons of clarity we use antisymmetrized interactions, i.e., $\tilde{V}_{\text{NN}}=\mathscr{A}_{\text{NN}} V_{\text{NN}}$ and 
$\tilde{V}_{\text{3N}}=\mathscr{A}_{\text{3N}} V_{\text{3N}}$, with antisymmetrization operators $\mathscr{A}_{\text{NN}}=1-P_{12}$ 
and $\mathscr{A}_{\text{3N}}=(1-P_{12})(1-P_{13}-P_{23})$. Up to second order in $\tilde{V}_{\text{NN}}$ and first order in 
$\tilde{V}_{\text{3N}}$ there are then four different contributions\footnote[2]{The second-order anomalous derivative term is not counted 
here; it follows immediately from $\Omega_0$ and $\Omega_{\text{1,NN}}$.}, represented diagrammatically in Fig.\ \ref{fig_mbds}.
%
\vspace*{1.1cm}
\subparagraph{Two-nucleon force.}
The first-order and the second-order normal contribution to the grand canonical potential density are given by 
\begin{align}
\label{diagramexp1NN}
\Omega_{1,\text{NN}}(\mu_0,T)=&\frac{1}{2}\tr_{\sigma_1,\tau_1}\tr_{\sigma_2,\tau_2}  \!\! \int  \! \! \frac{\mathrm{d}^3k_1 }{(2\pi)^3} \!\! \int  \! \! \frac{\mathrm{d}^3k_2 }{(2\pi)^3}\, \,
n_{ k_1} n_{k_2}  \, \, \braket{  \boldsymbol 1 \boldsymbol 2 \left| (1-P_{12}) V_{\text{NN}}\right|\boldsymbol 1 \boldsymbol 2} ,
\\
\label{diagramexp2n}
\Omega_{2,\text{normal}}(\mu_0,T) =&-\frac{1}{8} \left(
\prod_{i=1}^4 \tr_{\sigma_i,\tau_i}
\int  \! \! \frac{\mathrm{d}^3k_i }{\left(2 \pi \right)^3} \right) \left(2 \pi \right)^3
\delta (\vec{k}_1 +\vec{k}_2-\vec{k}_3-\vec{k}_4) \nonumber \\ &\times
 \frac{n_{k_1}n_{k_2} \bar{n}_{k_3} \bar{n}_{k_4}-
\bar{n}_{k_1} \bar{n}_{k_2}n_{k_3}n_{k_4}}
{\varepsilon_3+\varepsilon_4-\varepsilon_1-\varepsilon_2} 
\left| \braket{  \boldsymbol 1 \boldsymbol 2 \left| (1-P_{12})  V_{\mathrm{NN}}\right|\boldsymbol 3 \boldsymbol 4} \right|^2 ,
\end{align}
where $\bar n_k=1-n_k$.
The second-order anomalous contribution is given by
\begin{align}
\label{diagramexp2a}
\Omega_{2,\text{anomalous}}(\mu_0,T) =&-\frac{\beta}{2} \left(
\prod_{i=1}^3 \tr_{\sigma_i,\tau_i}
\int  \! \! \frac{\mathrm{d}^3k_i }{\left(2 \pi \right)^3} \right) 
n_{k_1}n_{k_2} \bar{n}_{k_2} n_{k_3} 
\nonumber \\ &\times
 \braket{  \boldsymbol 1 \boldsymbol 2 \left| (1-P_{12}) V_{\mathrm{NN}}\right|\boldsymbol 1 \boldsymbol 2} \,  \braket{  
 \boldsymbol 2 \boldsymbol 3 \left| (1-P_{12}) V_{\mathrm{NN}}\right|\boldsymbol 2 \boldsymbol 3}.
\end{align}

\begin{figure}[H] 
\vspace*{-2.9cm}
\hspace*{-0.55cm}
\includegraphics[width=1.1\textwidth]{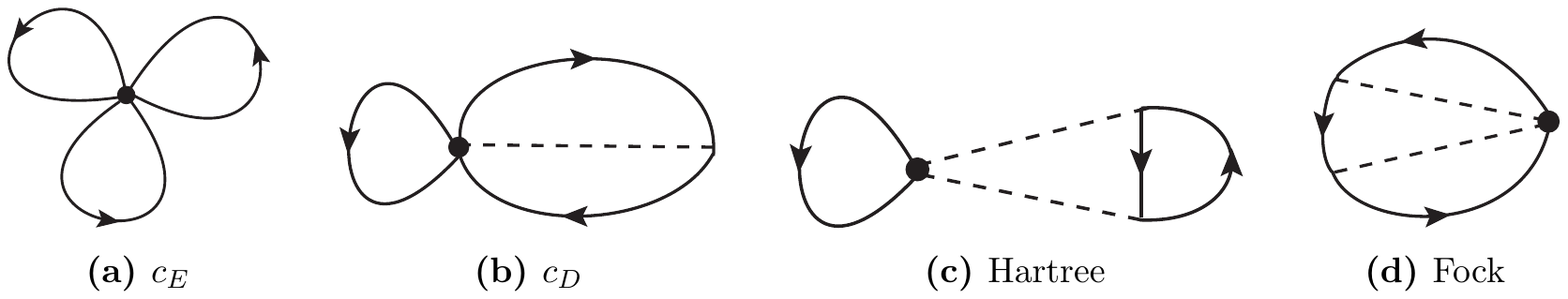} 
\vspace*{-19.73cm}
\caption{Contributions to $\Omega_{\text{1,3N}}$ from chiral 3N forces at N2LO. Dashed lines represent pions.}
\label{Figure3Ndiag}
\end{figure}

\subparagraph{Three-nucleon force.}
There are three different contributions to the N2LO three-nucleon force in $\chi$EFT: a three-nucleon contact diagram 
(proportional to the low-energy constant $c_E$), a one-pion exchange diagram (proportional to $c_D$), and a two-pion 
exchange diagram (with low-energy constants $c_1$, $c_3$ and $c_4$). The first-order many-body contribution arising 
from these diagrams can be written in the compact form
\begin{align}				\label{3N3bodyInt}
\Omega_{\text{1,3N}}(\mu_0,T)=
\int \limits_0^{\infty} \mathrm{d}k_1 \, \frac{k_1}{2 \pi^2} 
\int \limits_0^{\infty} \mathrm{d}k_2 \, \frac{k_2}{2 \pi^2}
\int \limits_0^{\infty} \mathrm{d}k_3 \, \frac{k_3}{2 \pi^2} \,\,
\mathcal{K}_3 \, n_{k_1} n_{k_2} n_{k_3} ,
\end{align}
where $\mathcal{K}_3=\mathcal{K}_3^{(c_E)} +\mathcal{K}_3^{(c_D)} +\mathcal{K}_3^{(\text{Hartree})} +
\mathcal{K}_3^{(\text{Fock})}$. The many-body diagrams associated with these four different kernels 
$\mathcal{K}_3^{(i)}$ are depicted in Fig.\ \ref{Figure3Ndiag}.

\vspace*{0.1cm}
\indent
When used in combination with low-momentum two-nucleon interactions, the three-body potential is usually 
multiplied with a smooth regulator in terms of Jacobi momenta $P$ and $Q$ \cite{Hebeler:2010xb}. Concerning 
the first-order 3N contributions we have found the effect of this regulator to be negligible over the range of 
considered densities and temperatures. Leaving out the regulator, the kernels can be simplified to the following 
expressions (where $g_A\simeq 1.29$ is the axial-vector strength, $f_\uppi\simeq 92.4 \, \text{MeV}$ is the 
pion decay constant, $\Lambda_\chi\simeq 700 \,\text{MeV}$, and $m_\uppi \simeq 138 \, \text{MeV}$ is the 
average pion mass):

\begin{align}
\mathcal{K}_3^{(c_E)} = -\frac{12 c_E}{f_\uppi^4 \Lambda_\chi} k_1 k_2 k_3,
\end{align}
\begin{align}
\mathcal{K}_3^{(c_D)} = \frac{3 g_A  c_D}{f_\uppi^4 \Lambda_\chi} k_3 \left( k_1 k_2 - \frac{m_\uppi^2}{4} \ln \frac{m_\uppi^2
+\left(k_1+k_2 \right)^2}{m_\uppi^2+\left(k_1-k_2 \right)^2} \right),
\end{align}
\begin{align} \label{Khartree}
\mathcal{K}_3^{(\text{Hartree})} =& \frac{3 g_A^2}{f_\uppi^4} k_3 \left[ 2 \left(c_3 -c_1 \right) m_\uppi^2 \ln \frac{m_\uppi^2
+\left(k_1+k_2 \right)^2}{m_\uppi^2+\left(k_1-k_2 \right)^2} - 4 c_3 k_1 k_2 \right. \nonumber \\
&\left. + \left(c_3 -2 c_1 \right)m_\uppi^4 \left(  \frac{1}{m_\uppi^2+\left(k_1+k_2\right)^2}-\frac{1}{m_\uppi^2+\left(k_1-k_2\right)^2} 
\right) \right],
\end{align}
\begin{align} \label{Kfock}
\mathcal{K}_3^{(\text{Fock})} = \frac{ g_A^2}{f_\uppi^4 k_3} \Big[ 3 c_1 m_\uppi^2 H(k_1) H(k_2) + \left( \frac{c_3}{2}-c_4 \right) 
X(k_1) X(k_2) + \left(c_3 +c_4 \right) Y(k_1) Y(k_2) \Big] .
\end{align}
\noindent
The functions $H(k_{i})$, $X(k_{i})$ and $Y(k_{i})$ in the Fock-contribution are:
\begin{align} 
H(k_i)&=k_i+\frac{k_3^2-k_i^2-m_\uppi^2}{4 k_3} \ln \frac{m_\uppi^2+\left(k_i+k_3 \right)^2}{m_\uppi^2+\left(k_i-k_3 \right)^2}, \\ 
X(k_i)&=2 k_i k_3-\frac{m_\uppi^2}{2} \ln \frac{m_\uppi^2+\left(k_i+k_3 \right)^2}{m_\uppi^2+\left(k_i-k_3 \right)^2}, \\ 
Y(k_i)&=\frac{k_i}{4 k_3} \left( 5 k_3^2 - 3 k_i^2 - 3 m_\uppi^2 \right) + \frac{3 \left( k_i^2-k_3^2+m_\uppi^2 \right)^2 + 4 m_\uppi^2 
k_3 ^2}{16 k_3^2} \ln \frac{m_\uppi^2+\left(k_i+k_3 \right)^2}{m_\uppi^2+\left(k_i-k_3 \right)^2} .
\end{align}

\subsection{Partial-wave representation of two-body contributions}\label{sec24}
The two-nucleon interaction $\tilde{V}_{\text{NN}}$ is usually given in terms of partial-wave matrix elements. In this section we 
give the partial-wave expanded form of the two-body contributions presented in the previous section, both for zero and for 
finite temperatures, as well as the expressions needed to calculate the second-order anomalous derivative term.

\subparagraph{First-order contribution.}
The partial-wave expansion of the matrix elements in Eq.\ (\ref{diagramexp1NN}) results in the following expression:
\begin{align}		\label{OmegaNN1pw}
\Omega_{1,\text{NN}}(\mu_0,T)=\frac{2}{\pi^3}\!  \int \limits_0^{\infty} \! \mathrm{d}p \, p^2 \int  \limits_0^{\infty}\!\! \mathrm{d}K \, K^2  
\mathcal{F}(p,K) 
\sum \limits_{J,\ell,S} (2J+1)(2\mathcal{T}+1)\,\, \braket{ p | \tilde{V}_{\text{NN}}^{J,\ell,\ell,S,\mathcal{T}}  | p } .
\end{align}
Here, $\tilde{V}_{\text{NN}}^{J,\ell_1,\ell_2,S,\mathcal{T}} $ are the matrix elements of the antisymmetrized two-body potential with 
respect to total angular momentum states $\Ket{J \ell_i S \mathcal{T}}$, and $\vec K$ is half the total momentum of the two nucleons. 
The function $\mathcal{F}(p,K) $ is given by
\begin{align} \label{thetaint1NN}
\mathcal{F}(p,K)=\int \limits_{-1}^{1} \mathrm{d}\cos \theta_K \, \, n_{\left|\vec{K}-\vec{p}\,\right|} n_{\left|\vec{K}+\vec{p}\,\right|}
=\frac{\ln(1+\e^{\eta+2x})-\ln \left(\e^{2x}+\e^{\eta}\right)}{x \left (\e^{2\eta}-1 \right)} \;,
\end{align}
where $\theta_K$ is the angle between $\vec K$ and $\vec p$, $x=\beta\frac{Kp}{2M}$ and $\eta =\beta 
\left( \frac{K^2+p^2}{2M} -\mu_0 \right)$.

The expression for the corresponding zero-temperature contribution to the energy per nucleon $\bar E$ can be simplified further. 
It is given by
\begin{align} \label{1storderNN}
\bar E_{1,\text{NN}}(k_F)=\frac{2}{\pi}\!  \int \limits_0^{k_{\!F}} \!\! \mathrm{d}p \, p^2  \left(1-\frac{3p}{2 k_F}+\frac{p^3}{2 k_F^3} 
\right)  
\sum \limits_{J,\ell,S} (2J+1)(2\mathcal{T}+1)\,\, \braket{ p |\tilde{V}_{\text{NN}}^{J,\ell,\ell,S,\mathcal{T}}  | p } .
\end{align}

\subparagraph{Second-order normal contribution.}
The partial-wave representation of the second-order normal contribution is given by
\begin{align} 		\label{Omega2n}
\Omega_{2,\text{normal}}(\mu_0,T)=&-\frac{8}{\pi^2} M  
\int \limits_0^{\infty} \! \mathrm{d}p_1 \, p_1^2  \int \limits_{-1}^{1} \mathrm{d}\cos \theta_1 
\int \limits_0^{\infty} \! \mathrm{d}p_2 \, p_2^2  \int \limits_{-1}^{1} \mathrm{d}\cos \theta_2 
\int  \limits_0^{\infty}\!\! \mathrm{d}K \, K^2 \, \,
\frac{\mathcal{F}(p_1,p_2,K,\theta_1,\theta_2)}{p_2^2-p_1^2}  
\nonumber \\  & \times  
\sum \limits_{S} \sum \limits_{J,\ell_1,\ell_2} \sum \limits_{J',\ell_1',\ell_2'} i^{\ell_2-\ell_1}
i^{\ell_1'-\ell_2'}
\braket{ p_1 | \tilde{V}_{\text{NN}}^{J,\ell_1,\ell_2,S,\mathcal{T}}  | p_2 }  \braket{ p_2 |\tilde{V}_{\text{NN}}^{J',\ell_2',\ell_1',S,\mathcal{T}} | p_1 } 
\nonumber \\  & \times   (2\mathcal{T}+1) 
\sum \limits_{M,m,m^\prime}  \,  {\cal C}(\theta_1,\theta_2) .
\end{align}
The function ${\cal C}(\theta_1,\theta_2)$ collects spherical harmonics and Clebsch-Gordan coefficients: 

\begin{align} \label{Xifunction}
{\cal C}(\theta_1,\theta_2)    =&
\mathcal{Y}_{\ell_1,(M-m)}(\theta_1)  \mathcal{Y}_{\ell_2,(M-m^\prime)}(\theta_2)  \mathcal{Y}_{\ell_2',(M-m^\prime)}
(\theta_2)  \mathcal{Y}_{\ell_1',(M-m)}(\theta_1) 
\nonumber \\  & \times  \braket{ \ell_1 (M-m) S m | J M \ell_1 S  }
\braket{ J M \ell_2 S | \ell_2 (M-m^\prime) S m^\prime} 
\nonumber \\ &  \times
\braket{ \ell_2' (M-m^\prime) S m^\prime | J' M \ell_2' S  }
\braket{ J' M \ell_1' S | \ell_1' (M-m) S m } .
\end{align}
Here, $\mathcal{Y}_{\ell, m}(\theta)$ denotes the spherical harmonics without the azimuthal part $ \e^{i m \phi}$.
The other function $\mathcal{F}(p_1,p_2,K,\theta_1,\theta_2)$ is given by
\begin{align}   	\label{Ffunction}
\mathcal{F}(p_1,p_2,K,\theta_1,\theta_2) =
n_{\left|\vec{K}+\vec{p}_1\right|} n_{\left|\vec{K}-\vec{p}_1\right|} \bar{n}_{\left|\vec{K}+\vec{p}_2\right|} \bar{n}_{\left|\vec{K}-\vec{p}_2\right|}
- \bar{n}_{\left|\vec{K}+\vec{p}_1\right|} \bar{n}_{\left|\vec{K}-\vec{p}_1\right|} n_{\left|\vec{K}+\vec{p}_2\right|} n_{\left|\vec{K}-\vec{p}_2\right|} ,
\end{align}
where the angles $\theta_{1,2}$ are measured with respect to (half) the total momentum $\vec K$.
Note that the integrand in Eq.\ (\ref{Omega2n}) is non-singular at $p_1=p_2$.

To obtain the zero-temperature expression for the second-order normal contribution the Fermi-Dirac distributions in Eq.\ 
(\ref{Ffunction}) have to be substituted with Heavyside step functions. The inequalities associated with these step functions can be 
absorbed into the boundaries of the integrals, which then results in the following expression:
\begin{align} \label{e2nTT0}
\bar{E}_{2,\text{normal}}(k_F) =&-\frac{24}{k_F^3} M 
\int \limits_{0}^{k_F} \!\! \mathrm{d}K \, K^2 \!\!\!\!
\int \limits_0^{\sqrt{k_F^2-K^2} }\!\!\!\!\!\!\! \mathrm{d}p_1 \, p_1^2   \!\!\!\!
\int \limits_{\sqrt{k_F^2-K^2} }^{\infty}\!\!\!\!\!\!\! \mathrm{d}p_2 \, p_2^2 \; \!\!\!\!
\int \limits_{-\text{min}( \alpha_1\,,\,1)}^{\text{min}( \alpha_1\,,\,1)} \!\! \!\!\!\!\!\!\mathrm{d}\cos \theta_1  \!\!\!\!
\int \limits_{-\text{min}( -\alpha_2\,,\,1)}^{\text{min}( -\alpha_2\,,\,1)} \!\! \!\!\!\!\!\!\mathrm{d}\cos \theta_2 \,\,
\frac{1}{p_2^2-p_1^2}\nonumber \\ \nonumber & \times
\sum \limits_{S} \sum \limits_{J,\ell_1,\ell_2} \sum \limits_{J',\ell_1',\ell_2'} i^{\ell_2-\ell_1}
i^{\ell_1'-\ell_2'}
\braket{ p_1 | \tilde{V}_{\text{NN}}^{J,\ell_1,\ell_2,S,\mathcal{T}}  | p_2 }  \braket{ p_2 |\tilde{V}_{\text{NN}}^{J',\ell_2',\ell_1',S,\mathcal{T}} | p_1 } 
\\  & \times  (2\mathcal{T}+1) 
\sum \limits_{M,m,m^\prime}    \,   {\cal C}(\theta_1,\theta_2) ,
\end{align}
where $\alpha_i=(k_F^2-K^2-p_i^2)/(2 K p_i)$.

\subparagraph{Second-order anomalous contribution.} 
Expanding the matrix elements in Eq.\ (\ref{diagramexp2a}) in terms of partial waves one arrives at
\begin{align}	\label{Omega2aPW}
\begin{split}
\Omega_{2,\text{anomalous}}(\mu_0,T) =&-\frac{16}{\pi^2} \beta 
\int \limits_0^{\infty} \!\! \mathrm{d}k \, k^2  n_k \bar{n}_k 
\left[  \int \limits_0^{\infty} \!\! \mathrm{d}p \, p^2    \sum \limits_{J,\ell,\ell',S} i^{\ell-\ell'}
\Braket{ \frac{p}{2} | \tilde{V}_{\text{NN}}^{J,\ell,\ell',S,\mathcal{T}}  | \frac{p}{2} }     \right.
\\ & \times \left.  \int \limits_{-1}^{1} \mathrm{d}\cos \theta_p \hspace{2mm}  n_{\left|\vec{p}+\vec{k}\right|}   \hspace{1mm}
 \sum_{M m_s t_z}\, {\cal C^\prime}(\theta_p)  \right] ^2  ,
\end{split}
\end{align}
where $\theta_p$ is the angle between $\vec p$ and $\vec k$, and
\begin{align}
{\cal C^\prime}(\theta_p)   =&
\mathcal{Y}_{\ell,(M-m_s)}(\theta_p)  \mathcal{Y}_{\ell',(M-m_s)}(\theta_p) 
\braket{ \ell (M-m_s) S m_s | J M \ell S  } \braket{ J M \ell' S | \ell' (M-m_s) S m_s } 
\nonumber \\  & \times 
\left|  \Braket{ S m_s | \left( m_s-1/2 \right) \, 1/2} \right|^2  
\left| \Braket{\mathcal{T} t_z |  \left( t_z-1/2 \right) \, 1/2  }\right|^2   .
\end{align}

\subparagraph{Second-order anomalous derivative term.} 
The numerator and denominator of the ADT contribution in Eq.\ (\ref{KLWseries}) can be evaluated 
separately. From Eq.\ (\ref{rhoformulae}) it follows immediately that
\begin{align} \label{ADT1aa}
\frac{\partial^2 \Omega_0(\mu_0,T)}{\partial \mu_0^2}=
-\frac{2M}{\pi^2} \int \limits_0^{\infty}\! \mathrm{d}p \, n_p ,
\end{align}
and from Eq. (\ref{OmegaNN1pw}) one gets
\begin{align}		\label{OmegaNN1pwderiv}
\frac{\Omega_{1,\text{NN}}(\mu_0,T)}{\partial \mu_0}
=\frac{2}{\pi^3}\!  \int \limits_0^{\infty} \! \mathrm{d}p \, p^2 \int  \limits_0^{\infty}\!\! \mathrm{d}K \, K^2  
\frac{\partial \mathcal{F}(p,K)}{\partial \mu_0}
\sum \limits_{J,\ell,S} (2J+1)(2\mathcal{T}+1)\,\, \braket{ p | \tilde{V}_{\text{NN}}^{J,\ell,\ell,S,\mathcal{T}}  | p } .
\end{align}
Here, the $\mu_0$ derivative of $\mathcal{F}(p,K)$ is given by
\begin{align}  \label{ADT1bb}
\frac{\partial \mathcal{F}(p,K)}{\partial \mu_0}=
\frac{\beta}{x} \left(
2\frac{\ln(1+\e^{\eta+2x})-\ln \left(\e^{2x}+\e^{\eta}\right)}
{\left(\e^{\eta}-\e^{-\eta} \right)^2}
+\frac{\e^{\eta} \left(1-\e^{4x} \right)}
{\left(\e^{2\eta}-1\right)\left(\e^{\eta}+\e^{2x} \right)\left(1+\e^{\eta+2x} \right)}
\right) ,
\end{align}
where $x$ and $\eta$ are the same as in Eq.\ (\ref{thetaint1NN}).
\\ \\ \indent
The numerical evaluation of the partial-wave representations of the second-order contributions at finite $T$ was tested with 
model
interactions of the one-boson exchange type \cite{HoltJW13a}. With such a simple form of the interactions a semi-analytical 
treatment
at second order is possible.
\subsection{Temperature- and density-dependent NN interaction}\label{sec25}
Up to now only the NN potential was considered in the second-order contributions. With three-nucleon forces included, the 
expressions for the second-order normal and anomalous contributions become somewhat involved. In zero-temperature 
many-body calculations it is common practice to approximate three-nucleon interactions at higher orders in perturbation
theory by using a density-dependent effective two-nucleon (DDNN) potential \cite{Holt:2008,Holt:2009ty,Hebeler:2009iv,
Hebeler:2010xb}. This potential is constructed from the genuine three-body force by integrating out one nucleon 
line by summing over occupied states in the Fermi sea. For details regarding the construction we refer to Ref.\ \cite{Holt:2009ty}. 
Generalizing to finite temperatures, the following replacements have to be made in Eqs.\ (11,12,17--25), of Sec. III A in 
Ref.\ \cite{Holt:2009ty}:
\begin{align} {k_f^3 \over 3} \rightarrow \int_0^\infty\!dk\,k^2 
\bigg[1+\exp{k^2/2M-
\mu_0\over T}\bigg]^{-1} = -\sqrt{\pi\over 2} (MT)^{3/2} {\rm Li}_{3/2}
(-e^{\mu_0/T}) = {\pi^2 \over 2} \rho(\mu_0,T).
\end{align}
The integrals over one pion propagator become
\begin{align} \Gamma_0(p) = {1\over 2p} \int_0^\infty\!dk\,k \bigg[1+
\exp{k^2/2M-\mu_0\over T}\bigg]^{-1}\ln{m_\uppi^2+(p+k)^2\over 
m_\uppi^2+(p-k)^2},
\end{align}
\begin{align} \Gamma_1(p) =  {1\over 4p^3} \int_0^\infty\!dk\,k\bigg\{4pk
- (m_\uppi^2+p^2+k^2) \ln{m_\uppi^2+(p+k)^2\over  m_\uppi^2+(p-k)^2} 
\bigg\}\bigg[1+
\exp{k^2/2M-\mu_0\over T}\bigg]^{-1}, 
\end{align}
\begin{eqnarray} \Gamma_2(p)&=& {1\over 16p^3} 
\int_0^\infty\!dk\,k\bigg\{4pk
(m_\uppi^2+p^2+k^2) -\Big[m_\uppi^2+(p+k)^2\Big] \Big[m_\uppi^2+(p-k)^2\Big]
\nonumber \\ && \qquad \qquad \qquad \times \ln{m_\uppi^2+(p+k)^2\over
m_\uppi^2+(p-k)^2} \bigg\}\bigg[1+\exp{k^2/2M-\mu_0\over T}\bigg]^{-1}, 
\end{eqnarray}
\begin{eqnarray} \Gamma_3(p)&=& {1\over 16p^5} 
\int_0^\infty\!dk\,k\bigg\{-12pk
(m_\uppi^2+p^2+k^2) +\Big[3(m_\uppi^2+p^2+k^2)^2-4p^2k^2\Big] \nonumber \\ &&
\qquad \qquad  \qquad\times \ln{m_\uppi^2+(p+k)^2\over m_\uppi^2+(p-k)^2} 
\bigg\}
\bigg[1+\exp{k^2/2M-\mu_0\over T}\bigg]^{-1},  \end{eqnarray}
while the integrals over the product of two different pion propagators are now given by
\begin{align} G_{0,*,**}(p,q) = {2\over q} \int_0^\infty\! dk\, 
{\{k,k^3,k^5\}
\over \sqrt{A(p)+q^2k^2} }\bigg[1+\exp{k^2/2M-\mu_0\over T}\bigg]^{-1}
\ln { q k+\sqrt{A(p)+q^2k^2}\over \sqrt{A(p)}} ,  
\end{align}
with $A(p)= [m_\uppi^2 +(p+k)^2][ m_\uppi^2+(p-k)^2]$. Note that Eqs.\ (19--22) in Ref.\ \cite{Holt:2009ty} which were 
set up to obtain the functions $G_{1,2,3}(p,q)$ remain valid.

\begin{figure}[H]
\vspace*{-1.0cm}
\hspace*{-0.1cm}
\includegraphics[width=1.05\textwidth]{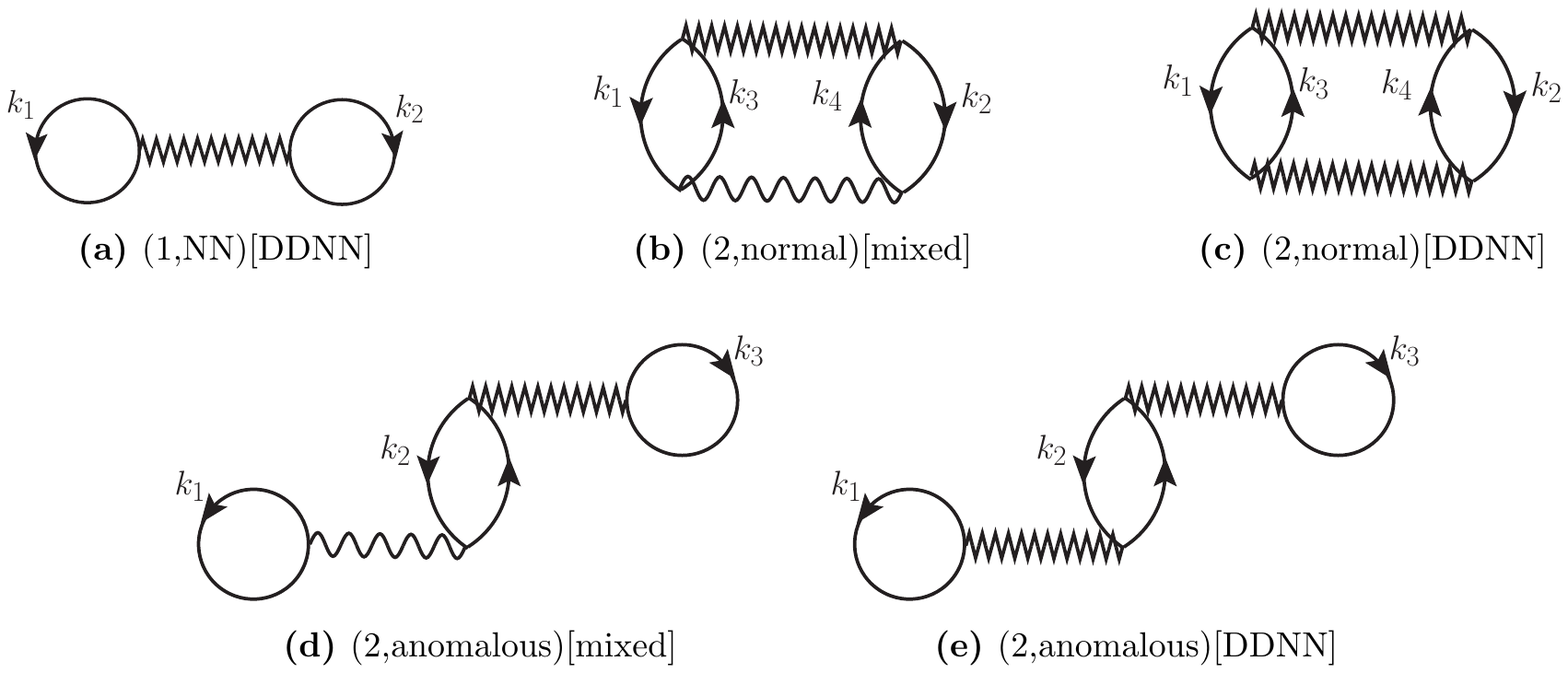} 
\vspace*{-15.4cm}
\caption{Antisymmetrized Goldstone diagrams representing the (a) first-order, (b-c)second-order normal, and (d-e) second-order
anomalous contributions associated with the DDNN potential (represented by zigzag lines). The NN potential is symbolized by wavy lines.
Diagram (a) carries an additional
symmetry factor of $1/3$, diagram (d) one of $1/2$, and diagram (e) one of $1/4$.}
\label{fig_mbdsDDNN}
\end{figure}

In this approximation three-nucleon forces are included at second order by substituting for $\tilde{V}_{\text{NN}}$ in Eqs.\ (\ref{Omega2n},\ref{e2nTT0},\ref{Omega2aPW},\ref{OmegaNN1pwderiv})
the quantity $\tilde V_{\text{DDNN}}(\rho,T)$ (the antisymmetrized temperature-dependent DDNN potential). When 
used in combination with $V_{\text{low-}k}(\Lambda)$ (i.e., in the VLK potential sets) the DDNN potential is constructed 
using the same sharp relative-momentum cutoff $\Lambda$, whereas when it is combined with the regularized chiral 
N3LO potentials (i.e., in the n3lo sets) the smooth regulator given in Eq.\ (\ref{EMregulator}) is used. The 
diagrammatic representations of the additional second-order normal and anomalous contributions arising from 
$\tilde V_{\text{DDNN}}(\rho,T)$ are depicted in Fig.\ \ref{fig_mbdsDDNN}.

%

\begin{figure}[H] 
\centering
\vspace*{-3.2cm}
\hspace*{-2.35cm}
\includegraphics[width=1.15\textwidth]{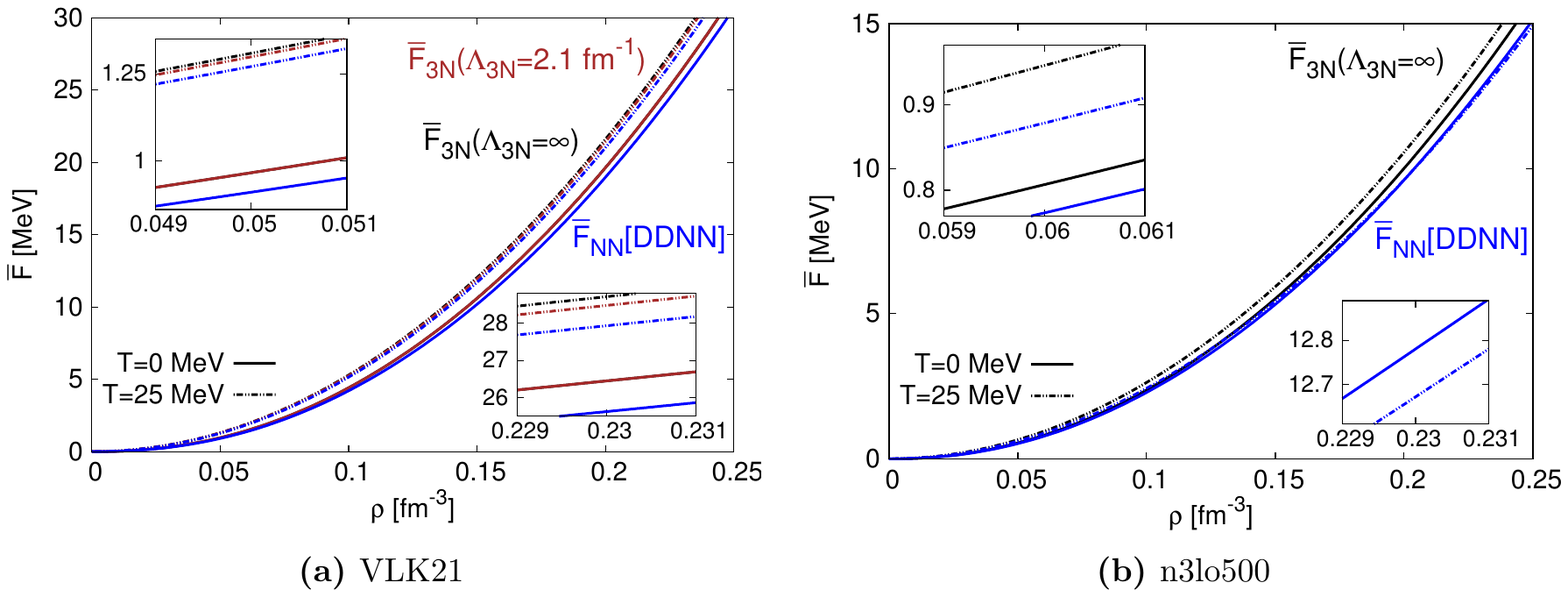} 
\vspace*{-17.0cm}
\caption{(Color online) First-order three-body contribution to the free energy per nucleon calculated with genuine 3N forces, $\bar F_{\text{3N}}$, and with the temperature- and density-dependent effective 
nucleon-nucleon (DDNN) potential, $\bar F_{\text{NN}}$[DDNN]. In Fig. (a) we show also the results when a Jacobi momentum 
regulator with $\Lambda_{\text{3N}}=2.1\,\text{fm}^{-1}$ is used for the genuine 3N contribution\protect\footnotemark. At zero temperature the results for $\bar{F}_{\text{3N}}(\Lambda_{\text{3N}}=\infty)$ and $\bar{F}_{\text{3N}}(\Lambda_{\text{3N}}=2.1\,\text{fm}^{-1})$ overlap.}
\label{DDNNplots3N}
\end{figure}

\newpage
\indent
To test the quality of the DDNN approximation we compare 
 in Figs.\ \ref{DDNNplots3N}(a) and \ref{DDNNplots3N}(b)
 the results for the first-order three-body contribution calculated 
with genuine 3N forces with the results obtained using $\tilde V_{\text{DDNN}}(\rho,T)$ in the first-order NN contribution, 
Eq.\ (\ref{1storderNN}). 
The 
quantity shown is the free energy per nucleon $\bar{F}(\rho,T)=\rho^{-1} F(\rho,T)$ as a function of density 
for temperatures $T=0, 25$ MeV. The insets
magnify 
the behavior in the low- and the high-density
region, respectively.
Incidentally one sees that 
the Nijmegen LECs used in the interaction sets involving $V_{\text{low-}k}$ potentials [Fig.\ \ref{DDNNplots3N}(a)] lead to considerably larger 3N 
contributions at first order in MBPT. For sharp regulators the T dependence of the first-order DDNN contributions is similar 
to the results obtained with genuine 3N forces. For the relatively soft $n=2$ regulator 
this is not the case in the high-density region, as can be seen in Fig.\ \ref{DDNNplots3N}b where the results for 
n3lo500 are shown. Nevertheless, the deviations are in all cases small enough to justify using $\tilde V_{\text{DDNN}}(\rho,T)$ 
instead of the genuine three-body potential at second order. 
\footnotetext[3]{To be precise, the regulator used to calculate the brown curves is given by $f(a,b)=\exp[-(a^2+
\frac{3}{4}b^2)^2/ \Lambda_{\text{3N}}^4]$, where  $a=\frac{1}{2}|k_1-k_2|$ and $b=\frac{2}{3}|k_3-\frac{1}{2}(k_1+
k_2)|$. As this regulator is more restrictive than the usual one where $a$ and $b$ are given by absolute values of 
(proper) Jacobi momenta, i.e.,  $a=|\vec P|$ and $b=|\vec Q|$, the effects of the latter are even smaller.   }

\begin{figure}[H] 
\centering
\vspace*{-2.0cm}
\hspace*{-2.35cm}
\includegraphics[width=1.15\textwidth]{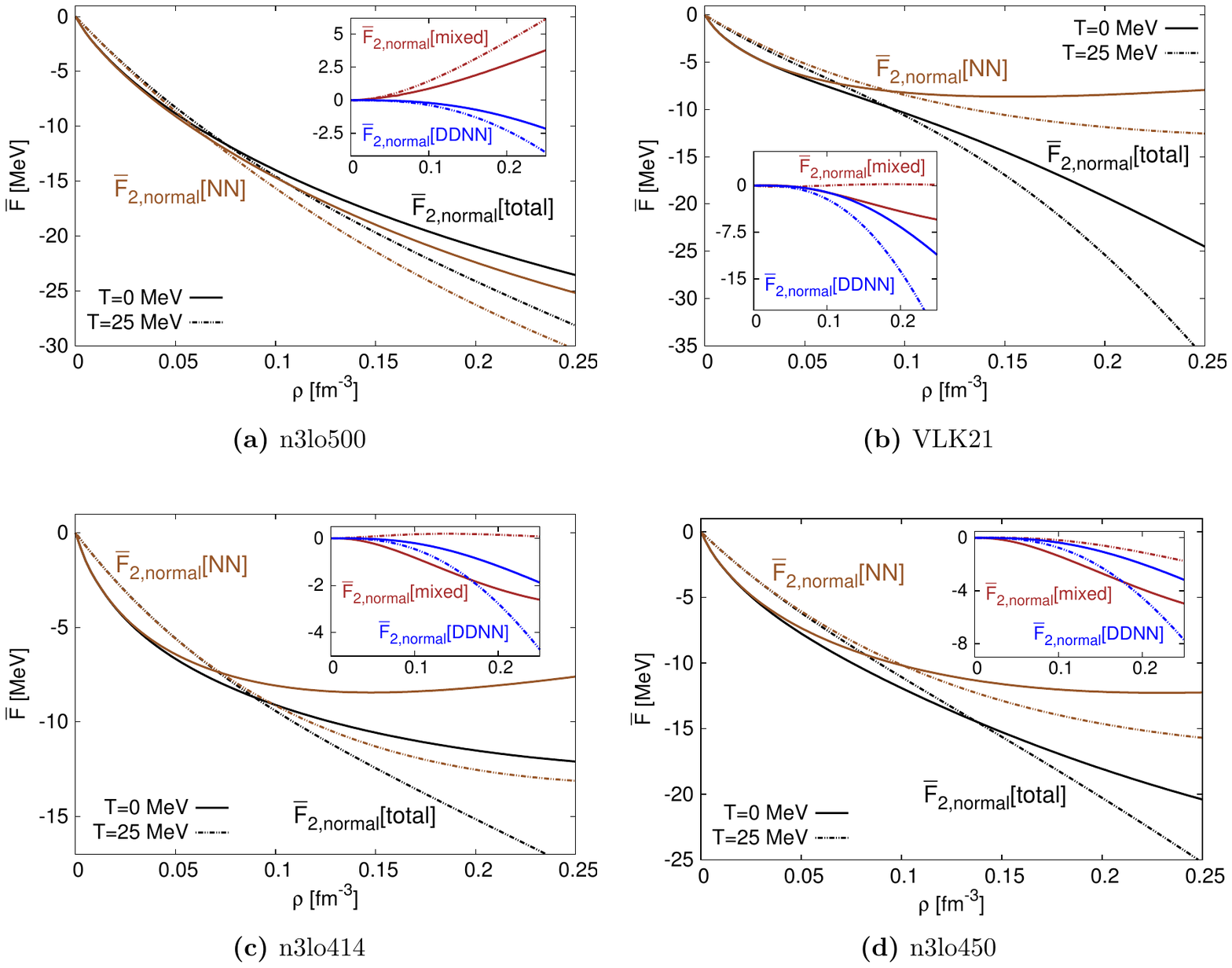} 
\vspace*{-10.8cm}
\caption{(Color online) Second-order normal contributions from different potential sets. The insets show the 
contributions which arise from the temperature- and density-dependent NN interactions.}
\label{DDNNplots2n}
\end{figure}

Figs.\ \ref{DDNNplots2n}(a) to \ref{DDNNplots2n}(d) show the results for the different second-order normal contribution arising 
from $\tilde{V}_{\text{NN}}$ and $\tilde V_{\text{DDNN}}(\rho,T)$. Here, $\bar F_{2,\text{normal}}[\text{NN}]$ denotes the 
contribution where both interactions are $\tilde{V}_{\text{NN}}$, $\bar F_{2,\text{normal}}[\text{mixed}]$ is the case where 
one interactions is given by $\tilde{V}_{\text{NN}}$ and other one by $\tilde V_{\text{DDNN}}(\rho,T)$, and $\bar 
F_{2,\text{normal}}[\text{DDNN}]$ denotes the case where both interactions are $\tilde{V}_{\text{DDNN}}(\rho,T)$. 
Furthermore, we define $\bar F_{2,\text{normal}}[\text{total}]=\bar F_{2,\text{normal}}[\text{NN}]+\bar F_{2,\text{normal}}[\text{mixed}]
 +\bar F_{2,\text{normal}}[\text{DDNN}]
 $.

One sees that the size of the two-body contribution $\bar F_{2,\text{normal}}[\text{NN}]$ increases with the 
resolution scale. Among the different NN potentials, n3lo500 gives rise to the largest second-order normal contribution. 
For n3lo450 and VLK23 (not shown) as well as for n3lo414 and VLK21 the results for $\bar F_{2,\text{normal}}[\text{NN}]$ 
are almost the same and feature non-monotonic behavior as the density and temperature increase. In contrast, similar 
to the first-order 3N contributions, the pure DDNN contributions $\bar F_{2,\text{normal}}[\text{DDNN}]$ exhibit a 
continuous increase in magnitude with density as well as with temperature. The size of the $\bar F_{2,\text{normal}}[\text{DDNN}]$ 
contribution is noticeably larger for VLK21 (and for VLK23). The size of the sum of the total additional DDNN contributions, 
$\bar F_{2,\text{normal}}[\text{mixed}]
 +\bar F_{2,\text{normal}}[\text{DDNN}]$, is then also the largest in that case. In the other cases 
$\bar F_{2,\text{normal}}[\text{DDNN}]$ is of comparable size. In the case of n3lo500 it is additionally suppressed by the mixed 
contribution $\bar F_{2,\text{normal}}[\text{mixed}]$, leading to an overall relatively small modification of the second-order normal contribution when three-body forces are included.
The different sizes of the additional DDNN contributions at second order for different potentials 
underlie most of the discussion in Sec. \ref{sec31}.

\vspace*{3.7cm}
\subsection{Results for anomalous contributions}  \label{sec26}
\begin{figure}[H] 
\centering
\vspace*{-2.2cm}
\hspace*{-2.35cm}
\includegraphics[width=1.15\textwidth]{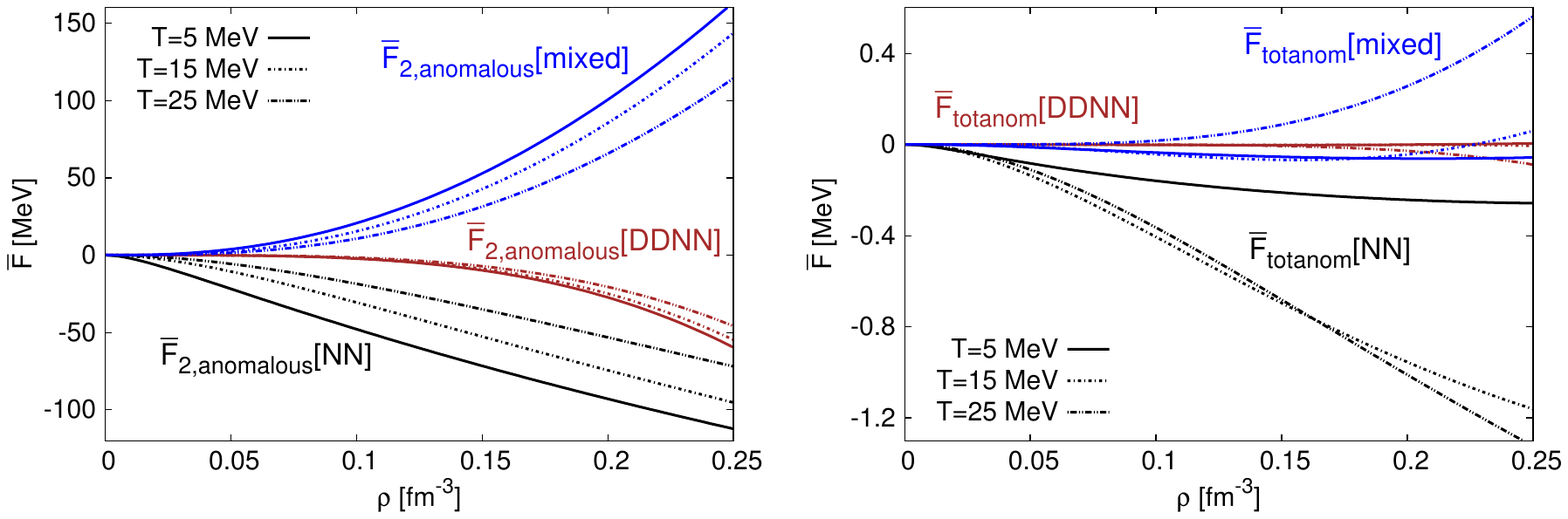} 
\vspace*{-17.9cm}
\caption{(Color online) Second-order anomalous contributions arising from $\tilde{V}_{\text{NN}}$ and $\tilde V_{\text{DDNN}}(\rho,T)$, and the 
corresponding contributions when the respective anomalous derivative terms are added, calculated using n3lo450.}
\label{ANOMplots}
\end{figure}

With $\tilde V_{\text{DDNN}}(\rho,T)$ included there are three different second-order anomalous contributions. 
The one where both interactions are $\tilde{V}_{\text{NN}}$ is denoted by $\bar F_{\text{2,anomalous}}[\text{NN}]$, the one 
with two $\tilde V_{\text{DDNN}}(\rho,T)$ type interactions is denoted by $\bar F_{\text{2,anomalous}}[\text{DDNN}]$, and
the case where one interaction is given by $\tilde{V}_{\text{NN}}$ and the other one by $\tilde V_{\text{DDNN}}(\rho,T)$ 
is denoted by $\bar F_{\text{2,anomalous}}[\text{mixed}]$. As can be seen in Fig.\ \ref{ANOMplots}, the size of these 
contributions is relatively large; in fact, in the high-density domain these are, together with the respective anomalous 
derivative terms, the largest contributions in the Kohn-Luttinger-Ward formula, Eq.\ (\ref{KLWseries}). However, the total 
anomalous contributions, i.e., $\bar F_{\text{totanom}}[\ldots]=\bar F_{\text{2,anomalous}}[\ldots]+\Bar F_{\text{ADT}}[\ldots]$, are relatively small in size and (as expected) decrease with temperature.

\vspace*{1.2cm}
\subsection{Self-energy contributions to the single-nucleon energies}  \label{sec27}
\begin{figure}[H] 
\centering
\vspace*{-3cm}
\hspace*{-2.2cm}
\includegraphics[width=1.15\textwidth]{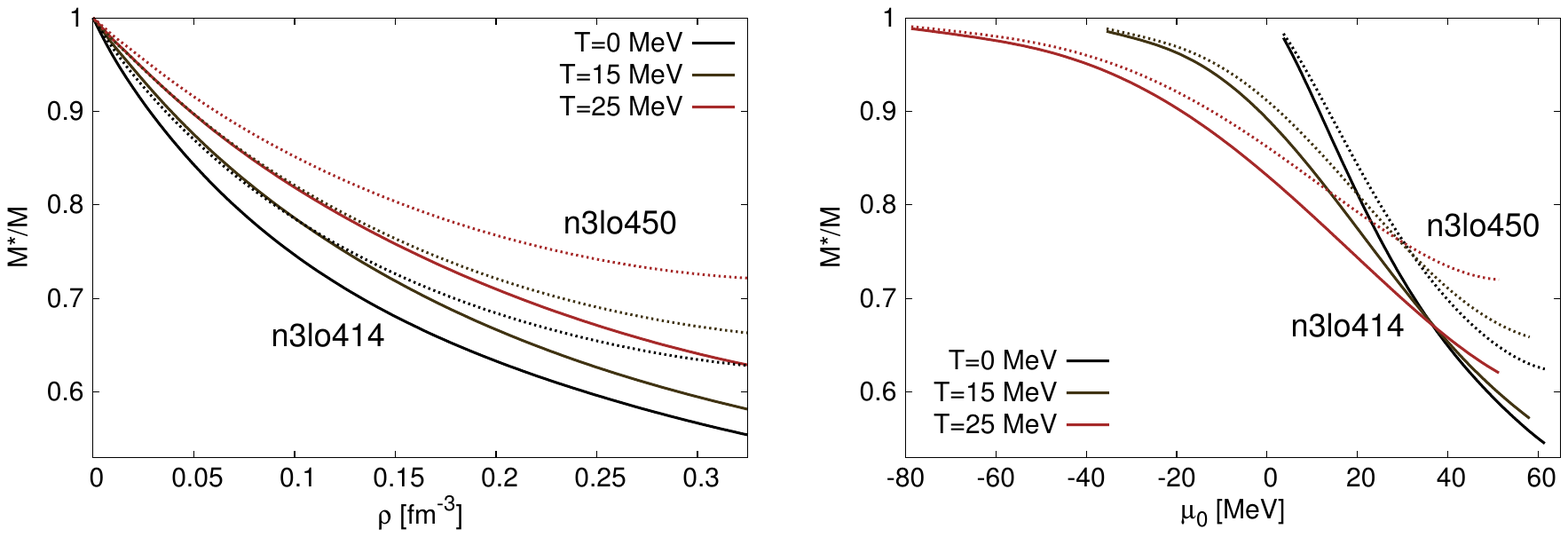} 
\vspace*{-17.9cm}
\caption{(Color online) Effective-mass ratio $M^{*}/M$ from n3lo414 (solid lines) and n3lo450 (dashed lines) as a function of the nucleon density $\rho$ and the one-body chemical potential $\mu_0$, respectively, for different temperatures.}
\label{Mstarplots}
\end{figure}

The single-nucleon energies $\varepsilon_k$ appear explicitly in the second-order normal contributions, Eq.\ (\ref{diagramexp2n}).  
The exact propagator $G(k,\omega)$ is defined by the self-consistent Dyson equation, which can be resummed as a geometric series:
\begin{align}  \label{Dyson Equation}
 G(k, \omega)=G_0(k, \omega)+G_0(k, \omega)\, \Sigma(k,\omega) \,G(k,\omega)=
 \left[ \omega-\frac{k^2}{2M}-\Sigma(k,\omega) \right]^{-1} .
\end{align}
Using the temperature-dependent DDNN potential approximation for three-body forces, the first-order contribution to the 
proper self-energy (expanded in partial waves) reads
\begin{align}		\label{selfenergyPW}
 \Sigma_1(k;\mu_0,T)=& \frac{1}{4\pi}\! \int \limits_0^{\infty} \!\! \mathrm{d}q \, q^2 \, n_{q} \! \int  \limits_{-1}^{1}\!\! \mathrm{d}\cos \theta_q  
\sum \limits_{J,\ell,S} (2J+1)(2\mathcal{T}+1)
 \Braket{ \frac{\abs{\vec k-\vec{q}\,}}{2} | 
 \tilde{V}_{\text{NN}}^{J,\ell,\ell,S,\mathcal{T}}+ \frac{1}{2}
 \tilde{V}_{\text{DDNN}}^{\,J,\ell,\ell,S,\mathcal{T}}
  | \frac{\abs{\vec k-\vec{q}\,}}{2} } ,
\end{align}
where $\theta_q$ is the angle between $\vec q$ and $\vec k$. The DDNN interaction carries an additional symmetry factor of $1/2$. Nucleon self-energies can be easily included using the 
effective-mass approximation:
\begin{align}	\label{effmassEq}
\varepsilon(k; \rho,T) = \frac{k^2}{2M} + \Sigma(k;\rho,T) \simeq \frac{k^2}{2M^{*}(\rho,T)}+U_0(\rho,T) ,
\end{align}
where $M^{*}(\rho,T)$ is called the (density- and temperature-dependent) effective mass. The momentum independent 
parts $U_0(\rho,T)$ of the single-nucleon energies cancel in Eq.\ (\ref{diagramexp2n}), and therefore it suffices to multiply 
the partial-wave expanded expressions in Eqs.\ (\ref{Omega2n}) and (\ref{e2nTT0}) with a factor $M^{*}(\rho,T)/M$ to incorporate self-energy effects. We show the size of this effective-mass factor for different temperatures and 
interactions in Fig.\ \ref{Mstarplots}. One sees that $M^{*}(\rho,T)/M$ decreases with density and increases with temperature
and that $M^{*}/M \leq 1$. Hence, including the effective-mass factors leads to a reduction of the different second-order 
normal contributions. 
When represented as a function of the one-body chemical potential $\mu_0$ the $M^{*}(\mu_0,T)/M$ curves all cross at approximately 
the same point
for each set of two- and three-body potentials, 
which is not directly apparent from Eqs. (\ref{selfenergyPW}) and (\ref{effmassEq}). Higher-order contributions to the nucleon
single-particle energies at or near zero temperature have been calculated from chiral nuclear interactions in Refs.\ 
\cite{Hebeler:2009iv,Holt11,Holt13a,Carbone2013}. Extending these calculations to the temperature region $T\leq 25$ MeV considered
in this work will be the subject of future research.

\vspace*{0.7cm}
\section{Results} \label{sec3}
In this section we examine the results for the thermodynamic equation of state of isospin-symmetric nuclear matter, 
calculated using all of the different contributions presented in the previous section. The convergence behavior of the 
many-body perturbation series and the model-dependence of the different contributions is investigated, and we then 
examine the (physical) equations of state resulting from the n3lo500, n3lo450 and n3lo414 potential sets.

\subsection{Convergence of the perturbation series} \label{sec31}
\begin{figure}[H] 
\centering
\vspace*{-2.9cm}
\hspace*{-2.2cm}
\includegraphics[width=1.15\textwidth]{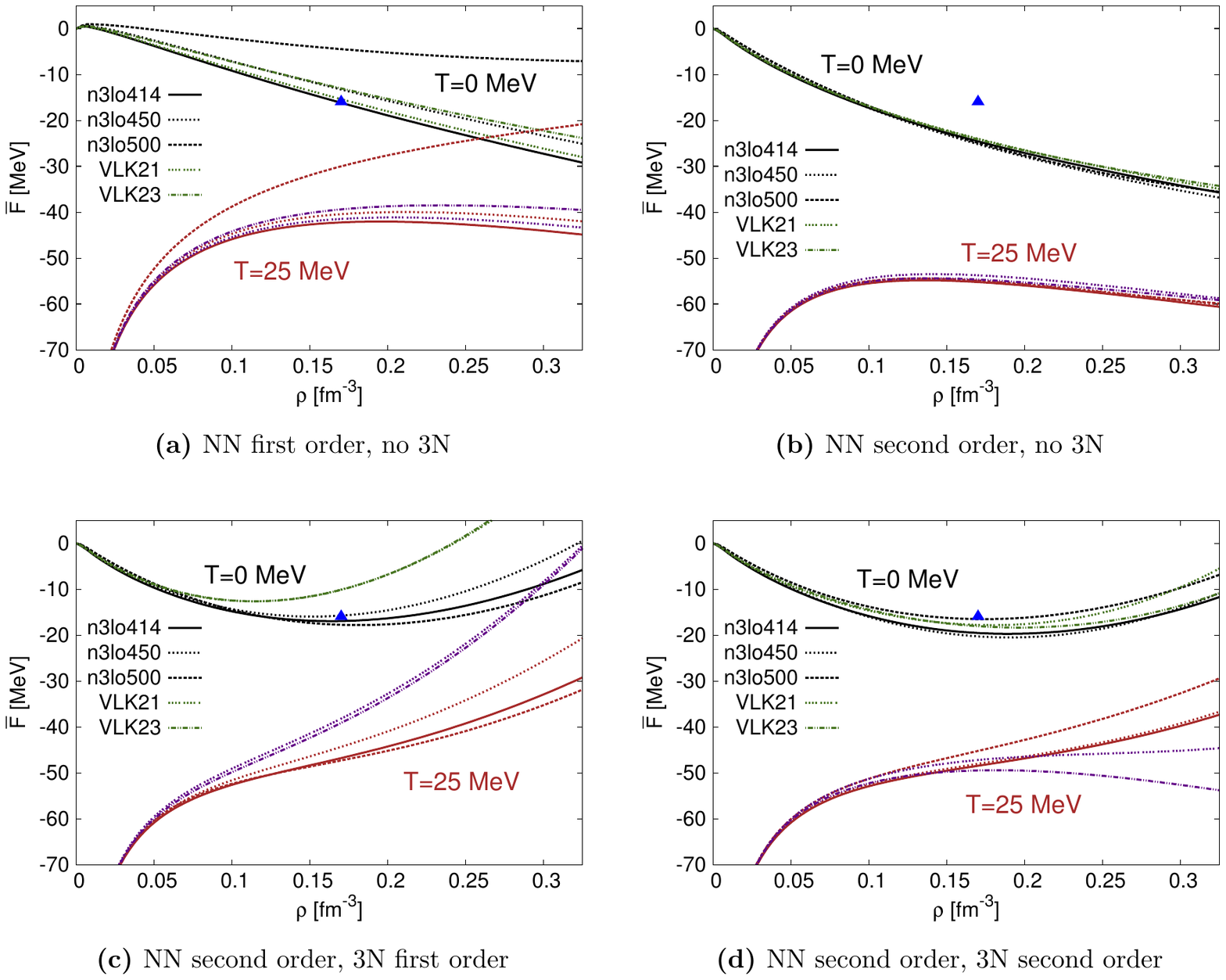} 
\vspace*{-10.7cm}
\caption{(Color online) Free energy per nucleon $\bar F(\rho,T)$ at different stages in MBPT, calculated using different low-momentum interactions. The 
blue triangle marks the empirical saturation point $\bar E_0\simeq -16\,\text{MeV}, \rho_0\simeq 0.17 \,\text{fm}^{-3}$. }
\label{Convergence4}
\end{figure}

In Fig.\ \ref{Convergence4} we examine the convergence behavior of the zero- and finite-temperature perturbation 
series for the different potential sets listed in Table \ref{table:LECs}. Figs. \ref{Convergence4}(a) and \ref{Convergence4}(b) 
show the results for the free energy per nucleon $\bar F(\rho,T)$ for the case when only two-body forces are considered. While the 
first-order results from n3lo450 and VLK23 as well as the ones from n3lo414 and VLK21 are of similar size, the free energy
per nucleon calculated from n3lo500 is significantly smaller in magnitude. 
In fact, for n3lo500 the first- and the
second-order
NN contributions are of comparable size, 
which points to
the decreased pertubative quality of this two-body potential
\cite{Coraggio:2014nvaa}. Even so, at second order (with
NN forces only) the scale-dependence is significantly reduced, and similar results
are 
obtained with all five interactions sets. 

The results deviate again 
when three-nucleon forces are included at first order, 
as can be seen in Fig.\ \ref{Convergence4}c, which is entirely from the deviating values 
of the five low-energy constants that parametrize the 3N potential. The deviations among the n3lo results (and similarly 
among the VLK results) are visible, but significantly smaller than the difference between VLK and n3lo results 
(particularly the VLK results at zero-temperature are almost identical at this order). Finally, in Fig.\ \ref{Convergence4}d 
the second-order DDNN contributions are included. Here the results become again more model-independent, but only in 
the case of zero temperature. At finite $T$ the curves for $\bar F(\rho,T)$ are now considerably flatter in the case of VLK21 and 
especially VLK23 as compared to the n3lo results. The reason for this behavior is the different size of the total second-order normal three-body contribution $\bar{F}_{\text{2,normal}}[\text{mixed}]+\bar{F}_{\text{2,normal}}[\text{DDNN}]$ in each case, cf.\ Figs.\ \ref{DDNNplots2n}(a) to \ref{DDNNplots2n}(d). This contribution 
is much larger for the VLK potential sets. At zero temperature it balances the large first-order 3N contribution caused by 
the Nijmegen LECs, leading to results similar to those of the n3lo LECs. Because of the much more pronounced temperature-dependence of the second-order DDNN contributions (as compared to the first-order three-body contribution) there is
overcompensation at finite T, leading to the observed flattening of the $\bar F(\rho,T)$ curves with increasing temperature. A 
similar (but more moderate) flattening occurs also in the high-density domain of the results obtained from n3lo450 and n3lo414. It is entirely 
absent in the case of n3lo500, where the respective contribution is small (and has opposite sign). 
Because the pressure is defined as $P(\rho,T)=\rho^2 \partial \bar F(\rho,T)/\partial \rho$, the flattening present in the 
VLK results leads to crossing pressure isotherms. Ultimately, the origin of this behavior lies in the large values of the
Nijmegen LECs.

\begin{figure}[H] 
\centering
\vspace*{-2.9cm}
\hspace*{-2.35cm}
\includegraphics[width=1.15\textwidth]{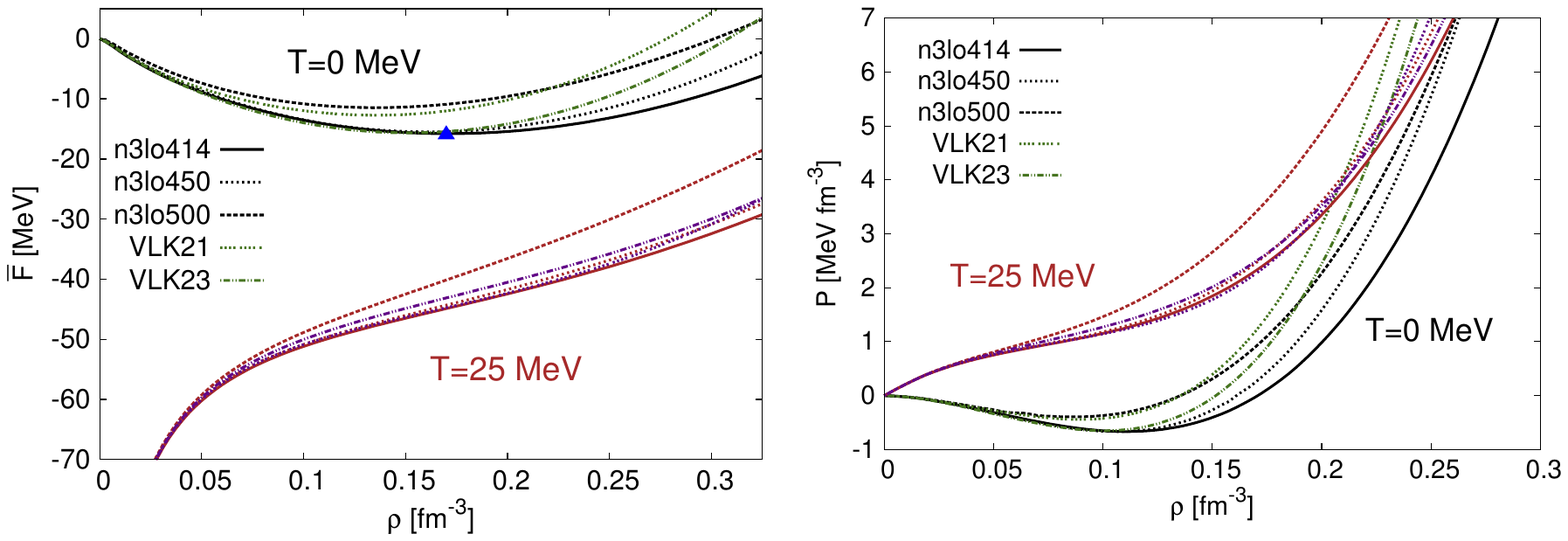} 
\vspace*{-17.9cm}
\caption{(Color online) Free energy per nucleon $\bar F(\rho,T)$ and pressure isotherms $P(\rho,T)$ for isospin-symmetric nuclear matter 
at second order in MBPT, with the effective-mass factors $M^{*}/M$ included. Only the pressure isotherms 
from VLK21 and the ones from VLK23 (green and purple curves) cross each other.}
\label{Convergence4Mstar}
\end{figure}

In Fig.\ \ref{Convergence4Mstar} we show the second-order results with the effective-mass factors $M^{*}/M$ 
included. The flatness problem of the finite temperature VLK curves is no longer present, and at $T=25\,\text{MeV}$ the 
VLK results and the ones obtained from n3lo450 and n3lo414 are in close agreement. At zero-temperature nuclear
matter is under-bound with the VLK21 potential ($\bar E_0=-12.73\,\text{MeV}$), the saturation density is somewhat 
small ($\rho_0=0.136\,\text{fm}^{-3}$), and the compressibility is $K=200 \,\text{MeV}$. For VLK23 the saturation point 
is close to the empirical value, i.e., $\bar E_0=-15.66\,\text{MeV}$ and $\rho_0=0.152\,\text{fm}^{-3}$, and 
the compressibility $K=260 \,\text{MeV}$ is in agreement with empirical constaints (see Sec. \ref{sec32}). However, for 
both VLK21 and VLK23 the zero-temperature curves are now somewhat steep for densities just above saturation density, and 
the crossing of the pressure isotherms is therefore still present as can be seen
in the second plot 
in Fig.\ \ref{Convergence4Mstar}\footnote[4]{A fully consistent RG treatment including induced many-nucleon forces may help 
cure this feature. See Refs.\ \cite{Bogner:2005sn,Jurgenson:2010wy,Hebeler:2012pr} for additional details.}. From $\partial P / 
\partial T=\upalpha/\upkappa_T$ (where $\upkappa_T 
\geq0$ is the isothermal compressibility) it follows that this crossing implies a large negative coefficient of thermal 
expansion $\upalpha$, i.e., there would be a large decrease in pressure when the temperature is increased at fixed 
density. In the case of n3lo450 and n3lo414 the pure second-order calculation resulted in nuclear matter that was
over-bound at low temperatures. The first-order corrections to the single-particle energies reduces the strong attraction
in the second-order normal diagram and improves the description of nuclear matter at zero temperature for the n3lo414
and n3lo450 potentials.
By contrast, with n3lo500 the saturation point is only reproduced in the pure second-order calculation (without the effective-mass corrections). 
The agreement is likely coincidental, and higher-order perturbative contributions should be included \cite{Coraggio:2014nvaa}.

%
%
\vspace*{1.9cm}
\subsection{Equation of state of isospin-symmetric nuclear matter} \label{sec32}
In the following we examine the properties of the nuclear matter equations of state that result from the chiral nuclear
interactions n3lo414, n3lo450 and n3lo500. Both the n3lo414 and n3lo450 potentials are well converged at second
order in perturbation theory with self-consistent single-particle energies \cite{Coraggio:2014nvaa}, while higher-order contributions to the free energy per particle are required to achieve convergence with the n3lo500 two-body 
potential. For comparison, we compute the thermodynamics of nuclear matter from n3lo500 {\it without 
effective-mass contributions} only to study whether universal features at finite temperature can arise starting from realistic 
zero-temperature equations of state.

\begin{figure}[H] 
\centering
\vspace*{1.6cm}
\hspace*{-2.3cm}
\includegraphics[width=1.15\textwidth]{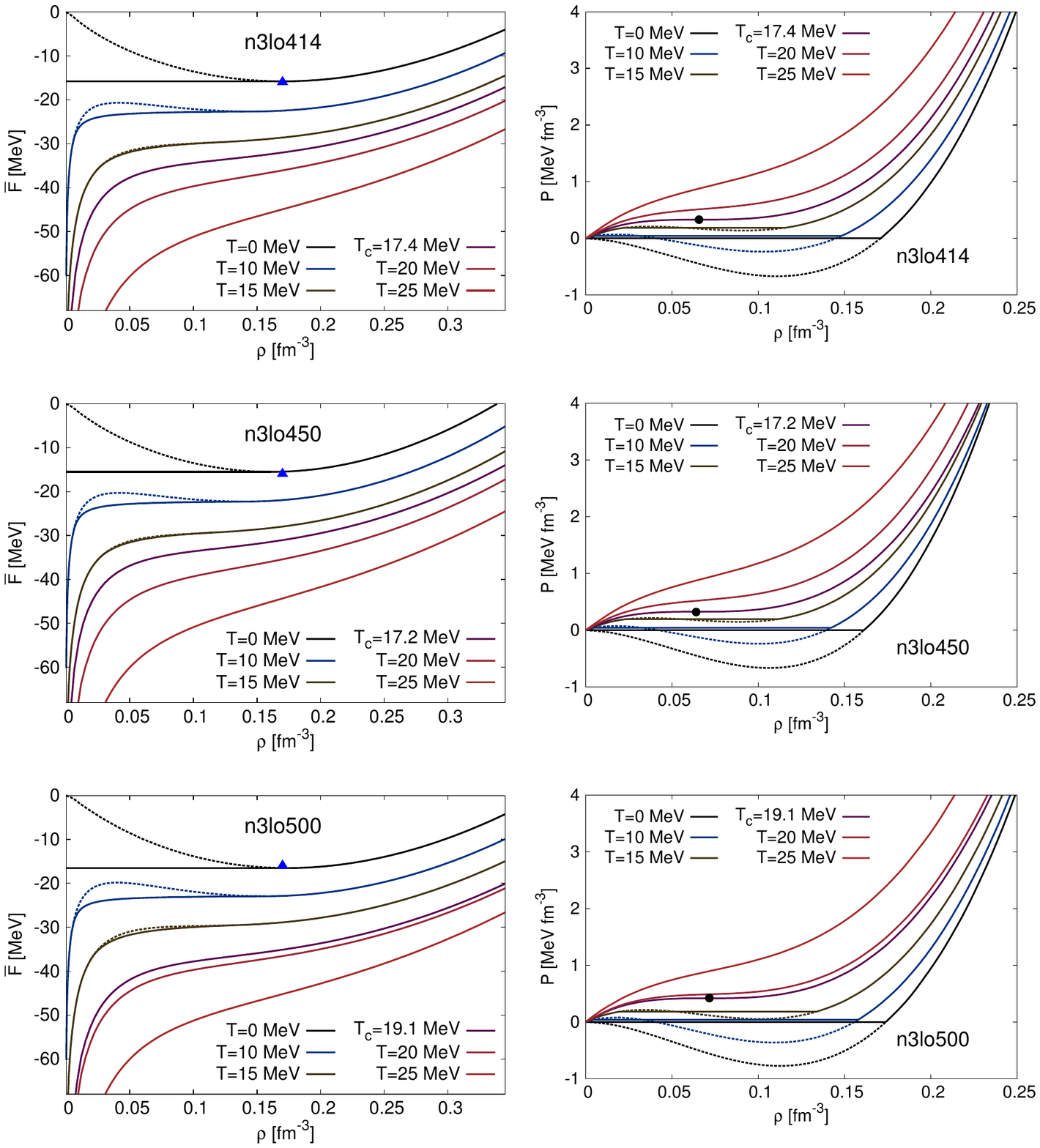} 
\vspace*{-8.3cm}
\caption{(Color online) Free energy per nucleon  $\bar F(\rho,T)$ and pressure $P(\rho,T$) for isospin-symmetric nuclear matter, 
calculated with the n3lo potential sets. The dashed lines show the analytical results, the solid lines the physical equations 
of state (obtained with the Maxwell construction). The blue triangle marks the empirical saturation point, and the black 
dot the critical point resulting from the respective potential sets.}
\label{EoSplots}
\end{figure}

\begin{figure}[H] 
\centering
\vspace*{-2.3cm}
\hspace*{-2.35cm}
\includegraphics[width=1.15\textwidth]{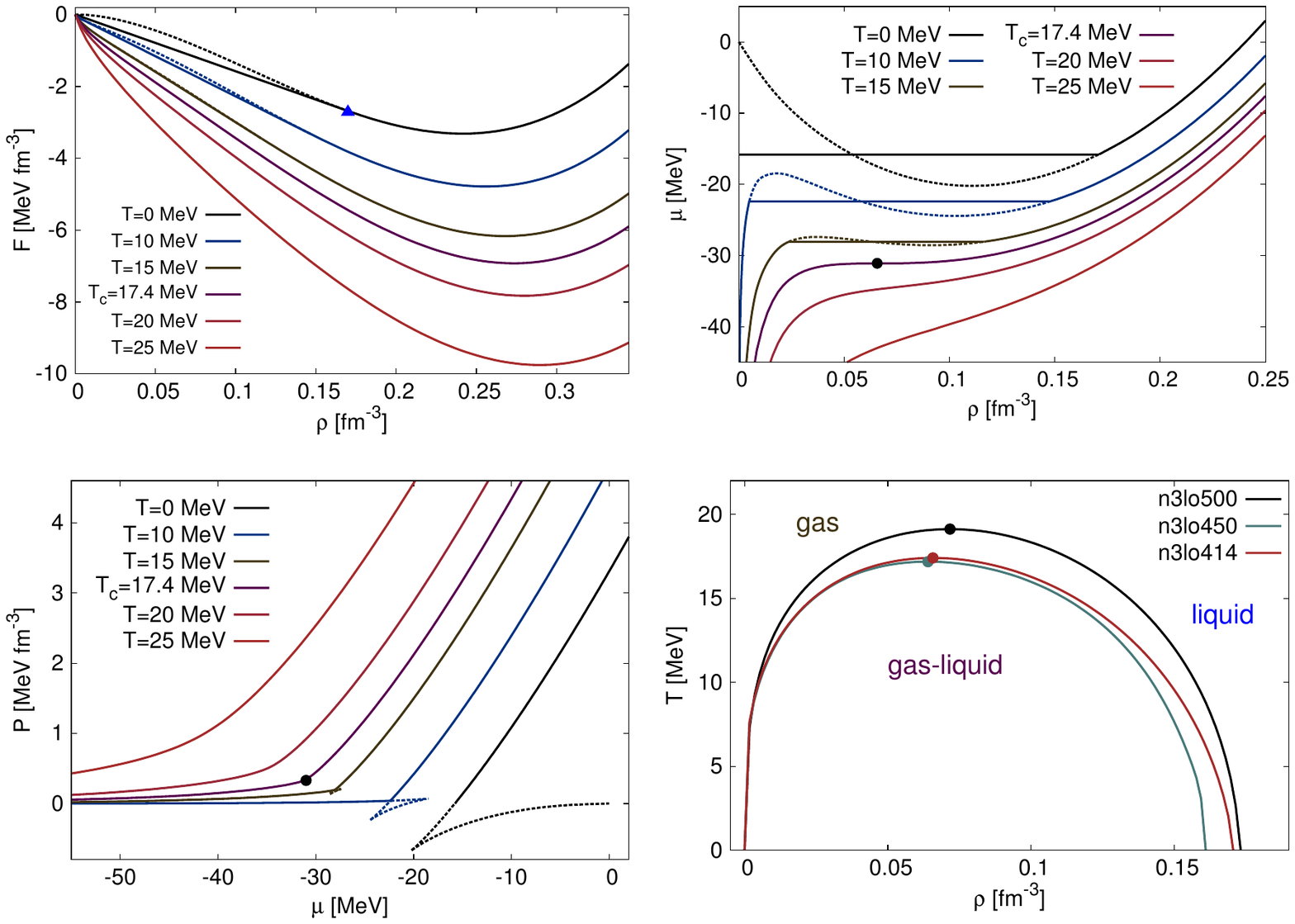} 
\vspace*{-11.5cm}
\caption{(Color online) Free energy density $F(\rho,T)$, chemical potential $\mu(\rho,T)$ and pressure as function of chemical 
potential $P(\mu,T)$ from n3lo414, as well as the $T-\rho$ phase diagram from n3lo500, n3lo450 and n3lo414. 
In the first three plots the dashed lines show the analytical results and the solid lines the physical equation 
of state.
The blue triangle and the (black) dot(s) mark the empirical saturation point and the 
determined critical point(s), respectively.}
\label{EoSplots_add}
\end{figure}

From the free energy per nucleon $\bar F(\rho,T)$ all other thermodynamic quantities follow by standard thermodynamic 
relations. The pressure $P(\rho,T)$ and the (nonrelativistic) chemical potential $\mu(\rho,T)$ for 
instance are given by
\begin{align}
P(\rho,T)=&\rho^2 \frac{\partial \bar{F}(\rho,T)}{\partial \rho} ,\\
\mu(\rho,T)=&\bar{F}(\rho,T)+\rho \frac{\partial \bar{F}(\rho,T)}{\partial \rho} .
\end{align}
The numerical results for $\bar F(\rho,T)$ and $P(\rho,T)$ are shown in Fig.\ \ref{EoSplots} for
densities $\rho<0.35\,\text{fm}^{-3}$ and temperatures in the region $T=0-25$ MeV
for all three n3lo potential sets. Additional derived thermodynamic quantities are shown for 
n3lo414 in Fig.\ 
\ref{EoSplots_add}, i.e., free energy density $F(\rho,T)$, chemical potential 
$\mu(\rho,T)$, and pressure $P(\mu,T)$. 

\subparagraph{Liquid-gas phase transition.}
For temperatures below a critical value $T_c$ the analytical free energy density at fixed temperature $F(\rho,T)=\rho 
\bar F(\rho,T)$ exhibits a mechanically unstable region of negative curvature\footnote[5]{Nonconvexity of $F(\rho,T)$ with respect to $\rho$ 
implies a negative isothermal compressibility $\upkappa_T$, which violates the stability relation $\upkappa_T \geq 0$.}, 
which signifies the presence of a first-order phase transition. The physical equation of state 
inside the transition region is obtained by performing the Maxwell construction. In the following we briefly recall the 
properties of the equation of state associated with this method.

%
%
%
%
%
%
%
%
%
%
%
%
\newpage

\indent
From the nonconvexity of the free energy density $F(\rho,T)$ it follows that the chemical potential and pressure isotherms 
are nonmonotonic (as functions of $\rho$). This implies that for temperatures $T<T_c$ there exist points $(\rho_a(T), F_a(T))$ and 
$(\rho_b(T), F_b(T))$ which have matching values of these quantities [denoted by $\mu_m(T)$ and 
$P_m(T)$] and therefore represent systems that can coexist in mutual thermodynamic equilibrium. 
These points delineate the region in which the low density, the gaslike (in analogy to classical gases described 
by the van der Waals equation of state), and the high density, the liquidlike phase, mix. 

In the regions adjacent 
to $(\rho_a, F_a)$ and $(\rho_b, F_b)$ where the free energy density is still convex the system is mechanically 
metastable, i.e., a finite disturbance is needed to induce phase separation, whereas in the inner region of 
thermodynamic instability the system separates spontaneously.

The values of $\mu_m(T)$ and $P_m(T)$ are obtained by constructing double tangents in the $\bar F(\nu,T)$ plots 
(where $\nu=1/\rho$ is the volume per nucleon), i.e., for fixed temperature $T<T_c$ one finds values $\nu_a$ and $\nu_b$ 
(where $\nu_a>\nu_b$) for which
\begin{align} \label{maxwell1}
\bar F(\nu_a,T)-\bar F(\nu_b,T)&=-P_m(T) (\nu_a-\nu_b), \\
\frac{\partial \bar F(\nu,T)}{\partial \nu} \Big|_{\nu_a, \nu_b}&=-P_{m}(T). \label{maxwell2}
\end{align}
The points specified by these equations are identical to the ones with equal values of pressure and chemical potential. The 
free energy per nucleon of the liquid-gas mixture is then given by substituting the analytical results with the 
double tangents,\footnote[6]{The concentrations of the liquid and the gas part in the phase separated system are given by 
$c_{\text{liquid}}(\nu)=\frac{\nu-\nu_b}{\nu_a-\nu_b}$ and $c_{\text{gas}}(\nu)=\frac{\nu_a-\nu}{\nu_a-\nu_b}$, respectively, 
so the free energy per nucleon of the mixture coincides with the one given by the double tangents, $c_{\text{liquid}}(\nu) \bar 
F(\nu_a,T)+c_{\text{gas}}(\nu) \bar F(\nu_b,T)=\bar F(\nu_b,T)-P_{m}(T) (\nu-\nu_b)$. Note that negative curvature of $F(\rho,T)$ corresponds to concavity of $\bar F(\nu,T)$ (at fixed $T$), so the 
double tangents lie underneath the analytical results and the free energy density of the mixture is smaller than that of the 
unseparated system.} i.e., for $\rho \in [\rho_a(T), \rho_b(T)]$ 
and $T<T_c$ it is
\begin{align}
\bar{F}(\rho,T)=\mu_{m}(T)-\frac{P_{m}(T)}{\rho} .
\end{align}
The physical equations of state resulting from this construction are given by the solid lines in Figs.\ \ref{EoSplots} and \ref{EoSplots_add}. Since the Maxwell construction does not preserve the 
curvature of $\bar F(\rho,T)$ at the boundaries $\{\rho_a(T),\rho_b(T)\}$ of the transition regions, both $P(\rho,T)$ and 
$\mu(\rho,T)$ are not differentiable at these points. For $\rho \in \left[\rho_a(T),\rho_b(T)\right]$ the chemical potential and 
the pressure are constant and their values given by $\mu_{m}(T)$ and $P_{m}(T)$, respectively. Hence, in the physical 
$P(\mu,T)$ diagrams the regions of phase coexistence collapse to single points with coordinates $(P_{m}(T),\,\mu_{m}(T))$. 
The different parts of the region of thermodynamic instability are particularly exposed in the analytical $P(\mu,T)$ curves. 
Here, the transition from mechanical metastability to the unstable region with nonconvex free energy density is marked out by 
sharp bends and for $T<T_c$ the analytical $P(\mu,T)$ diagrams become triple-valued (double valued at zero temperature). The regions of 
phase coexistence terminate at the critical point $(P_c, \rho_c, T_c)$ where both derivatives of the pressure vanish (signifying a
 second-order transition point):
\begin{align}
\frac{\partial P(\rho,T)}{\partial \rho}\Big|_{T=T_c,\, \rho=\rho_c}=
\frac{\partial^2 P(\rho,T)}{\partial \rho^2}\Big|_{T=T_c,\, \rho=\rho_c}=0 .
\end{align}
For pressures above the critical value $P_c$ there is no phase transition from a dense liquidlike to a low density gaslike phase; 
nuclear matter instead behaves as a fluid whose properties vary continuously with temperature.

\subparagraph{Zero-density limit.}
In the limit of vanishing density the interactions between nucleons vanish and  $\bar F_0=\mu_0+ \Omega_0/\rho$ gives the 
dominant contribution to the free energy per nucleon. The singular behavior of the $\bar F(\rho,T)$ curves for $\rho \rightarrow 0$ at 
non-zero temperature is therefore entirely caused by the non-interacting contribution $\bar F_0$. The leading term in Eq.\ 
(\ref{OmegaNorbert}) can be written as
\begin{align}
\Omega_0(\mu_0,T)=\sqrt{2}\,T\left(\frac{M}{\beta \pi}\right)^{\frac{3}{2}} 
\text{Li}_{5/2}\big(-\exp(\beta \mu_0)\big) \,.
\end{align}
With the corresponding expression for $\rho(\mu_0,T)$ in Eq. (\ref{rhoformulae}) and $\mu_0 \xrightarrow{\rho \rightarrow 0} 
-\infty$ it follows that
\begin{align}
\bar F_0(\mu_0,T)=\mu_0-T \frac{\text{Li}_{5/2}\big(-\exp(\beta \mu_0)\big)}{\text{Li}_{3/2}\big(-\exp(\beta \mu_0)\big)} 
\xrightarrow{\mu_0 \rightarrow -\infty}\mu_0-T .
\end{align}
Returning to Eq.\ (\ref{mufreelimit}) this shows that the singularity of the free energy per nucleon is logarithmic,  $\sim \ln \rho$.
Hence, despite 
the divergent behaviour of $\bar F(\rho,T)$, the free energy density $F(\rho,T)=\rho \bar F(\rho,T)$ vanishes in the $\rho 
\rightarrow 0$ limit (cf.\ Fig.\ \ref{EoSplots_add}). Furthermore, the limiting behavior of the entropy per nucleon is 
given by
\begin{align}
\bar S (\rho,T)=-\frac{\partial \bar F(\rho,T)}{\partial T} \xrightarrow{\rho\rightarrow 0}
\frac{5}{2}-\frac{\mu_0(\rho,T)}{T}.
\end{align}
This shows that the total internal energy per nucleon $\bar E=\bar F +T\bar S$ approaches the value $3T/2$ for $\rho \rightarrow 0$, 
which corresponds to the equation of state of a classical ideal gas. 

It should be noted here that in the low density region the picture of nuclear matter as a homogeneous system is incomplete. 
This follows from the fact that at densities well below saturation density few-body correlations as well as Coulomb repulsion of protons are important. Light clusters 
such as deuterons, tritons and $\upalpha$ particles are formed. Because of the Pauli principle these clusters dissolve at higher 
densities, yet they can still be expected to play a role when it comes to the liquid-gas phase transition. A detailed study of cluster 
formation and its effects on the nuclear EoS was provided by Typel et al. \cite{Typel}. Their results suggest only modest 
changes regarding the position of the critical point, such as a shift of $T_c$ by less then $10\%$ from nucleonic clustering.

%
%
%
%
%
%
\newpage

\subparagraph{Nuclear bulk properties and thermodynamic observables.}
In Table \ref{table:observables} we give the values of several key quantities that characterize the obtained equations of state. 
The empirical saturation point $(\bar E_0, \, \rho_0)\simeq (-16 \, \text{MeV},\, 0.17\,\text{fm}^{-3})$ is best reproduced by n3lo414, 
but all potential sets lead to acceptable agreement. Also the empirical value of the compressibility \cite{Vretenar:2003qm, 
chen2009higher, Blaizot} 
\begin{align}
K = 9 \rho_0^2 \,\frac{\partial^2 \bar E(\rho)}{\partial \rho^2} \Big|_{\rho=\rho_0}=250 \pm 25\, \text{MeV}
\end{align}
comes out correctly for all sets of potentials. The critical point lies higher for n3lo500, and is very similar for the n3lo450 and 
n3lo414 results. The most recent empirical values for its coordinates have been obtained by the study of data from 
multifragmentation and compound nuclear decay experiments by Elliot \emph{et al.}\ \cite{Elliott:2013pna}; their values 
$T_c=17.9 \pm 0.4\,\text{MeV}$, $\rho_c=0.06 \pm 0.02\,\text{fm}^{-3}$, $P_c=0.31 \pm 0.07\,\text{MeV fm}^{-3}$ are in agreement 
with previous studies by Karnaukhov \emph{et al.}\ \cite{Karnaukhov:2008be}, and agree more closely with the n3lo450 and n3lo414 results.
\begin{table}[H]
\begin{center}
\begin{minipage}{14cm}
\setlength{\extrarowheight}{1.5pt}
\hspace*{5mm}
\begin{tabular}{|l|| c |c |c ||c |c |c|}
\hline 
& $\bar E_0\,[\text{MeV}]$ & $\rho_0\,[\text{fm}^{-3}]$ & $K\,[\text{MeV}]$ &$T_c\,[\text{MeV}]$&$\rho_c\,[\text{fm}^{-3}]$&$P_c
\,[\text{MeV fm}^{-3}]$
   \\ \hline 
   \hline
n3lo500 (no $M^{*}/M$) &-16.51   & 0.174  &250   & 19.1  &  0.072 & 0.42 \\
n3lo450 ($M^{*}/M$)  &-15.50   & 0.161  &244   &  17.2 &  0.064 & 0.32 \\
n3lo414 ($M^{*}/M$) &-15.79   & 0.171  & 223   &  17.4 &  0.066 & 0.33  \\  \hline 
\end{tabular}
\end{minipage}
{\vspace{0mm}}
\begin{minipage}{14cm}
\caption
{Saturation point $(\bar E_0,\rho_0)$, compressibility $K$, and critical values of temperature $T_c$, density $\rho_c$ and 
pressure $P_c$ resulting from the equations of state obtained with the n3lo potentials.}
\label{table:observables}
\end{minipage}
\end{center}
\end{table}
%
%
%
%
\subparagraph{Phase diagram.}
In Fig.\ \ref{EoSplots_add} we show the $T-\rho$ phase diagrams resulting from n3lo500, n3lo450 
and n3lo414. As a consequence of the third law of thermodynamics, the boundaries of the coexistence region 
$(\rho_a(T), \rho_b(T))$ must approach the $\rho$-axis with infinite slope. At zero temperature there is no pure gas 
phase, and the boundary points are given by  $(0, \rho_0)$. Above the critical temperature there is only the gaslike phase.

\subsection{Discussion of results}\label{sec34}
We have seen that the differences in the results obtained from different potential sets are predominantly from the 
contributions associated with the three-body interactions, which depend sensitively on the choice of low-energy constants 
$c_E$, $c_D$ and $c_{1,3,4}$. The dominant three-body contributions are the ones which are 
proportional to $c_3$, and the crossing of pressure isotherm present in the VLK21 and VLK23 results can be linked mainly to the 
large value of this low-energy constant in the Nijmegen LECs.

It should be stressed that in our calculation we have used leading-order (with respect to the chiral expansion) three-body forces 
only. The subleading (N3LO) 3N forces and the leading 4N forces have so far been fully included only in 
neutron matter calculations \cite{Tews:2012fj,Kruger:2013kua} at zero temperature. In nuclear matter already the leading-order 
3N force is more intricate; in addition to the vanishing of all contributions proportional to $c_E$, $c_D$ and $c_4$, in pure neutron 
matter the two-pion exchange kernels given in Eqs.\ 
(\ref{Khartree}) and (\ref{Kfock}) are decreased by factors $1/12$ and $1/6$, respectively\footnote[7]{These reduced isospin factors
follow from the absence of proton lines 
in the Hartree- and Fock-diagrams in Fig.\ \ref{Figure3Ndiag}.}. 
Initial investigations have shown that chiral four-body forces can give contributions to the nuclear equation of state of 
considerable size, but substantial cancellations among the contributions from N3LO many-nucleon forces have been 
conjectured \cite{Kaiser:2012ex}. It remains a future task to fully include higher-order many-nucleon forces in nuclear matter 
calculations.

\begin{figure}[H] 
\centering
\vspace*{-2.9cm}
\hspace*{-2.2cm}
\includegraphics[width=1.15\textwidth]{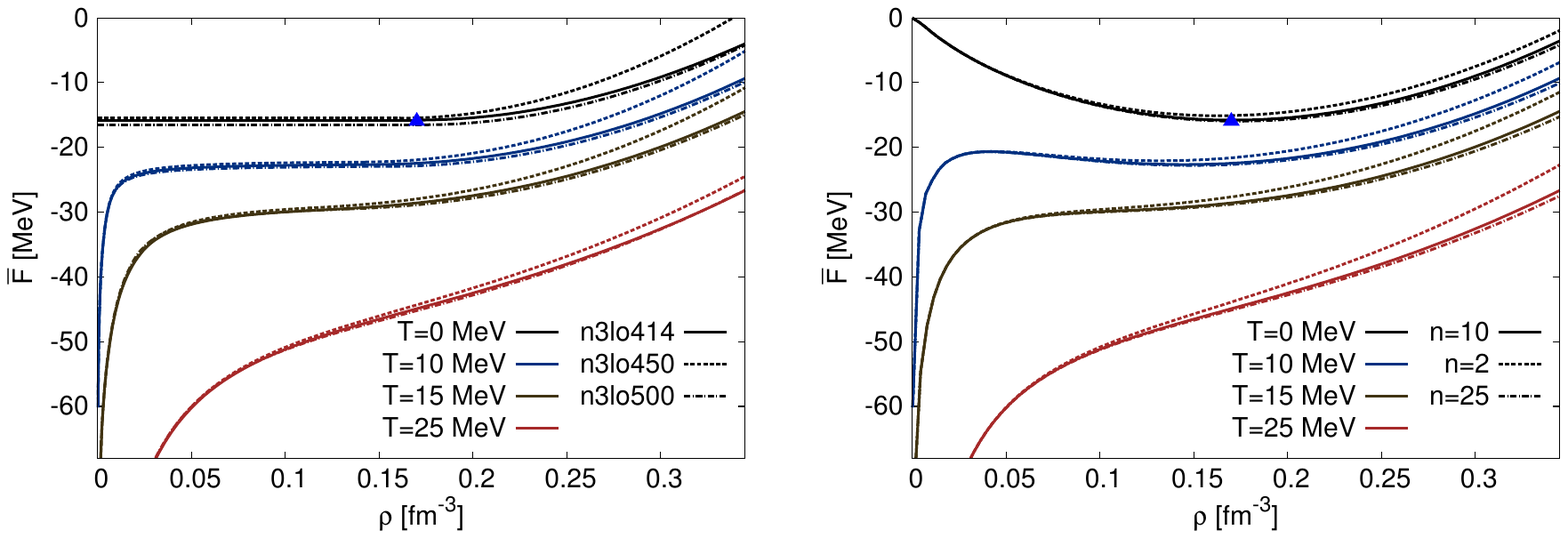} 
\vspace*{-17.9cm}
\caption{(Color online) Uncertainty bands in the results for the free energy per nucleon $\bar F(\rho,T)$ resulting from the different n3lo 
potential sets, i.e., from differently regularized chiral interactions, and from independently varying the DDNN regulator. The curves in the second plot have been calculated using n3lo414 and show the results without the Maxwell construction applied. The blue triangle marks the empirical saturation point.}
\label{finalplots}
\end{figure}

The potentials considered in the present work that best reproduce bulk properties of symmetric nuclear matter at zero 
temperature also give comparable results at finite temperature. In Fig.\ \ref{finalplots} uncertainty estimates
derived from variations in the cutoff scale and nuclear contact terms are shown. For densities $\rho \gtrsim 
\rho_0$ the deviations increase, and are (surprisingly) larger between n3lo450 and n3lo414 as compared to n3lo500 and 
n3lo414. Fig.\ \ref{finalplots} also shows the effect of varying the width of the DDNN regulator independently (with respect 
to the NN regulator). One sees that moderate variations have no large impact 
on the results. In particular, the effect is almost identical for different temperatures.
%

\vspace*{1.2cm}
\section{Summary} \label{sec5}
We have calculated the thermodynamic equation of state of isospin-symmetric nuclear matter using nuclear 
potentials derived within the framework of chiral effective field theory. The dependence of nuclear matter
properties on both the choice of the cutoff scale in the regulating function as well as the values of the 
low-energy constants associated with the N2LO chiral three-nucleon force were studied. Thermodynamically 
consistent results have been obtained with chiral nuclear potentials whose two-body low-energy constants 
have been fit to NN scattering phase shifts at the cutoff scales $414, 450, 500\,\text{MeV}$ and whose three-nucleon
contact terms were fit to the triton binding energy and lifetime. The results presented in this work (particularly
for the critical temperature, critical density, and critical pressure) therefore represent genuine predictions of 
nuclear many-body dynamics with constraints coming only from nuclear few-body systems. 
In the cases considered, 
good reproduction of the zero-temperature saturation point and compressibility led to consistent thermodynamics, 
and in particular a narrow range for the critical temperature 
$T_c = 17.2-19.1$ MeV
of the liquid-gas phase transition.

In future work we plan to extend our calculations to the case of isospin-asymmetric nuclear matter, with pure
neutron matter as a limiting case. This will allow for the comparison of additional observables such as the symmetry 
energy and the isobaric compressibility. Such calculations will be key to constructing microscopic equations of
state for use in numerically intense simulations of astrophysical phenomena. Additionally, from the quark-mass 
dependence of the chiral potentials it will be possible to determine the thermodynamic properties of the in-medium 
chiral condensate related to spontaneous symmetry breaking. Including the effects of subleading many-nucleon forces 
as well as explicit $\upDelta(1232)$-isobar degrees of freedom represent future challenges. 
\vspace*{1.2cm}
\acknowledgements
This work is supported in part by BMBF, by the DFG - NSFC (CRC 110), and US DOE Grant No. DE-FG02-97ER-41014.

\newpage
\bibliographystyle{apsrev4-1}	
\bibliography{refs}		
\end {document}